\documentclass[preprintnumbers,superscriptaddress,floatfix,prd]{revtex4}
	
		\usepackage{longtable}
	\usepackage{amsmath,amssymb,bm}
	\usepackage{graphicx}
	\usepackage{hyperref}
	\usepackage{amsfonts}
	\usepackage{amssymb}
	\usepackage[caption=false]{subfig}
	\usepackage{color}
	\newcommand{\beq}{\begin{eqnarray}}
	\newcommand{\eeq}{\end{eqnarray}}

\begin{document}

		\title{Hydrodynamic Excitations in Hot QCD Plasma }
		
		\author{Navid Abbasi}
		
		\affiliation{School of Particles and Accelerators, Institute for Research in Fundamental Sciences (IPM), P.O. Box 19395-5531, Tehran, Iran}

			\author{ Davood Allahbakhshi}
		
		\affiliation{School of Particles and Accelerators, Institute for Research in Fundamental Sciences (IPM), P.O. Box 19395-5531, Tehran, Iran}

		\author{ Ali Davody}
		\affiliation{Institute of Theoretical Physics, Regensburg University, 93040 Regensburg, Germany}

		\author{Seyed Farid Taghavi}
		
		\affiliation{School of Particles and Accelerators, Institute for Research in Fundamental Sciences (IPM), Tehran, Iran}
		
		\begin{abstract}
			{We study the long wavelength excitations in rotating QCD fluid  in presence of an  external magnetic field at finite vector and axial charge densities. 
				We consider the fluctuations of vector and axial charge currents coupled to energy and momentum fluctuations and compute the $SO(3)$ covariant dispersion relations of the six corresponding  hydrodynamic modes.  
				Among them, there are always two scalar Chiral Magnetic-Vortical-Heat (CMVH) waves;
				In the absence of magnetic field (vorticity) these waves reduce to CVH (CMH) waves.
				While CMVH waves are the mixed of CMH and CVH waves, they  have generally different velocities compared to the sum of velocities of the latter waves. The other four modes, which are made out of scalar-vector fluctuations, are mixed Sound-Alfv\'en waves. We show that when magnetic field is parallel with the vorticity, these four modes are the two ordinary sound modes  together with two Chiral Alfv\'en waves (CAW) propagating along the common direction of the magnetic field and vorticity. }
		\end{abstract}
		\maketitle

	\section{Introduction}
	\label{1}
	The phenomenon of chiral (or anomaly induced) transport was firstly studied in the  fermionic systems either weakly coupled to the electromagnetic field \cite{Vilenkin:1980fu} or in presence of rotation in the system \cite{Vilenkin:1979ui}. Due to presence of anomaly in the chiral system, macroscopic currents may be produced along the external magnetic field (Chiral Magnetic Effect, CME) or  along the vorticity of the system (Chiral Vortical Effect, CVE). The appearance of such currents is the main feature of the chiral transport theory which has been extensively studied in the literature. For example in the context of kinetic theory, a chiral theory  has been derived from the underlying quantum field theory \cite{Son:2012wh,Stephanov:2012ki} in which, the Berry
	monopole is responsible for the CME and CVE \cite{Stephanov:2012ki,Chen:2012ca}. Chiral magnetic effect  has been also studied numerically via lattice field theory \cite{Buividovich:2009wi,Buividovich:2010tn,Puhr:2016kzp,Buividovich:2016ulp}.

	In another direction, after the fluid-gravity duality showed the possibility of presence of the missed vorticity term \cite{Banerjee:2008th,Erdmenger:2008rm}, the issue of chiral transport was taken under study in hydrodynamics.  At first sight, the parity breaking terms like magnetic field and the vorticity seem to be in contradiction with the existence of a positive  divergence entropy current in fluid dynamics, however  the necessity of the second law of thermodynamics makes a relation between these terms of the hydrodynamic currents with the underlying quantum anomalies \cite{Son:2009tf}.

	In contrast to the coefficients of the dissipative transport, the coefficients of the parity odd terms may be entirely fixed in terms of both  the anomaly coefficients and the thermodynamic variables. The anomaly induced transport is in fact a non-dissipative phenomenon and so the associated coefficients are referred to as the so-called non-dissipative transport coefficients \cite{Kharzeev:2011ds}. This kind of hydrodynamic  transport has been used to effectively describe different phenomena in physics; e.g. in neutron stars or supernova in  astrophysics \cite{Sen:2016jzl,Yamamoto:2016zut,Kaminski:2014jda}, in the study of the origin of the magnetic fields in cosmology \cite{Giovannini:1997eg} and in propagation of Helicons in Weyl semi-metals in condensed matter physics\cite{Pellegrino:2014}. (See also \cite{Landsteiner:2013sja,Landsteiner:2016led}.)
	
	It has been argued that in a hot plasma of chiral fermions, e.g.  the  plasma of quarks and gluons produced in heavy ion collisions, the combination of the Chiral Separation Effect (CSE) \cite{Son:2004tq} and CME \cite{Fukushima:2008xe} gives rise to the propagation of a new kind of gapless excitation through the hot plasma; it is called the Chiral Magnetic Wave (CMW) \cite{Kharzeev:2010gd}.
	CMWs have been exploited to
	predict the charge asymmetries in the final state of a heavy ion collision \cite{Burnier:2011bf}. Consistent with the prediction of chiral transport, 
	the  charge asymmetries have been actually detected in experiments at RHIC and LHC \cite{Belmont:2014lta,Adamczyk:2015eqo}. The similar predictions have been made for the propagation of Chiral Vortical Wave (CVW) in heavy ion plasma in \cite{Jiang:2015cva}.

	The CMW found in \cite{Kharzeev:2010gd} has been computed by considering the fluctuations of vector and axial charge densities, keeping the local energy and  local momentum in the plasma fixed. The same result has been found in the context of chiral kinetic theory in   \cite{Stephanov:2014dma}.
	As discussed in  \cite{Stephanov:2014dma}, the assumption of getting the charge fluctuations decoupled from the energy-momentum fluctuations might be justifiable in high temperature low density regime  or even in high density low temperature regime for large $N$ theories.  In this paper we compute the spectrum of the hydrodynamic excitations in the most general case in which  the vector  and axial charges fluctuations are coupled to fluctuations of energy and momentum and compute the corresponding spectrum of the hydrodynamic excitations in magnetic field. We find the full spectrum of hydrodynamic modes
	in the system, including six different waves.
	As a result, in addition to the two scalar  Chiral Magnetic-Heat-Waves (CMHWs), we find another four collective excitations, each of them being a coherent perturbation of all six hydrodynamic variables. We will show that these four modes are mixed Sound-Alfv\'en waves. In the direction of magnetic field, the latter four modes are identified with two ordinary sound waves together with two Chiral-Alfv\'en-Waves (CAWs).  The propagation  of CAW was first predicted theoretically in a chiral fluid 
	with a single chirality \cite{Yamamoto:2015ria,Abbasi:2015saa} in the Landau-Lifshitz frame. It has been shown that the linear fluctuations of the vorticity may couple to the magnetic field and produce a wave of momentum perturbations propagating parallel to the magnetic field. This wave, namely CAW, might even propagate at zero density in the single chirality fluid. We will show that for the propagation of the CAW in QCD fluid, it is needed  both the vector and axial chemical potentials to be non-zero.

	We also repeat the above computations for the case of a rotating QCD type hot plasma. As a result we find two scalar Chiral-Vortical-Heat-Waves (CVHWs) together with four mixed Sound-Coriolis waves. In the direction of vorticity, the latter four modes are identified with two ordinary sound waves together with two rigid rotation modes.

	As a main part in the paper, we compute the hydrodynamic excitations in a rotating hot plasma which is simultaneously coupled to an external magnetic field. In this case the CMHW  and CVHW mix with each other and make Chiral-Magnetic-Vortical-Heat-Waves (CMVHWs).   It has been shown that in a fluid with turned off momentum perturbations the velocity of mixed waves might be equal to the sum of the velocities of individual waves when the magnetic field is parallel to the vorticity\cite{Chernodub:2015gxa,Frenklakh:2016izv}. However, we show that when taking into account the momentum perturbations, even for the vorticity being along the direction of the magnetic field, the velocity of mixed waves is not in general  equal to the sum of the velocities of the CMHW and CVHW.

	Let us emphasize that all of our results in the paper are $SO(3)$ covariant. This means that, not only propagation of hydro modes whether parallel or perpendicular to the magnetic filed and vorticity are studied here, but also our results are able to explain the propagation of waves in every another arbitrary direction with respect to the magnetic field and the rotation axis.
	
	The paper has been organized as follows. We begin with a brief review of the hot chiral QCD plasma in section \ref{section_chiral fluid}.  The content of sections \ref{section_B} and \ref{section_Omega} is related to detailed computations of magnetic field and vorticity respectively. In section \ref{section_mixed} we consider the general case wherein, both the magnetic field and the vorticity are present. In section \ref{pheno}, we use apply our theoretical results to the case of QGP. In the same section, we study the effect of dissipation on the hydro modes. We end with conclusion and  
	mentioning some follow up questions in the section \ref{conclu}.
	
	\section{Hydrodynamic of a QCD-type Fluid}
	\label{section_chiral fluid}
	We consider the QCD matter in an external magnetic field. In addition to usual electric charge (with vector current $J^{\mu}=\bar{\psi}\gamma^{\mu}\psi$), this matter carries the chiral charge (with chiral current $J^{\mu}_5=\bar{\psi}\gamma^{\mu}\gamma^{5}\psi$).
	In presence of a background gauge field $A_{\mu}$, 
	the dynamical equations  for this hot matter are nothing but the following conservation equations:
	\begin{equation}\label{EoM}
	\begin{split}
	\partial_{\mu}T^{\mu \nu}=&\,F^{\nu \lambda} J_{\lambda}\\
	\partial_{\mu} J^{\mu}=&\, 0\\
	\partial_{\mu} J_5^{\mu}=&\, \mathcal{C} E_{\mu} B^{\mu}
	\end{split}
	\end{equation}
	where $J^{\mu}=J_R^{\mu}+ J_L^{\mu}$ and $J_{5}^{\mu}=J_R^{\mu}- J_L^{\mu}$ and $\mathcal{C}$ is the chiral anomaly coefficient.
	In long wavelength regime, when the matter is in local equilibrium state, the energy momentum tensor $T^{\mu\nu}$, vector current $J^{\mu}$ and the chiral current  $J^{\mu}_5$ may all be effectively expressed in terms of six degrees of freedom: three components of the flowing matter velocity $u^{\mu}$, energy density $\epsilon$, vector charge density $n$ and axial charge density $n_5$.  We may also define the electric and magnetic field in the rest frame of this fluid  as $B^{\mu}=\frac{1}{2}\epsilon^{\mu\nu\alpha\beta}u_{\nu}F_{\alpha \beta}$ and  $E^{\mu}=\,F^{\mu \nu}u_{\nu}$, respectively \cite{Son:2009tf}.
	For small deviations from local equilibrium state,  each of the constitutive relations of the fluid may be given in a derivative expansion: 
	\begin{equation}\label{TJ}
	\begin{split}
	T^{\mu \nu}=& \,(\epsilon+p) u^{\mu} u^{\nu}+ p \,\eta^{\mu \nu} +\tau^{\mu \nu}\\
	J^{\mu}=& \,n u^{\mu} +\nu^{\mu},\\
	J_5^{\mu}=& \,n_5 u^{\mu} +\nu^{\mu}_5
	\end{split}
	\end{equation}
	with $\tau^{\mu \nu}$, $\nu^{\mu}$ and $\nu^{\mu}_5$ as the derivative corrections to fluid currents.
	In the Landau-Lifshitz frame  where $u_{\mu}\tau^{\mu \nu}=0$, $u_{\mu} \nu^{\mu}=0$ and $u_{\mu} \nu^{\mu}_5=0$ \cite{Landau}, up to first order in derivative expansion  we have 
	\begin{eqnarray}\label{disspart}
	\tau^{\mu \nu}&=&-\eta  P^{\mu\alpha}P^{\nu\beta}\left(\partial_{\alpha}u_{\beta}+\partial_{\beta}u_{\alpha} \right)-\left(\zeta-\frac{2}{3}\eta\right) P^{\mu \nu} \partial.u\\ \label{disspart_vector}
	\nu^{\mu}&=&  - \sigma T P^{\mu \nu} \partial_{\nu}\left(\frac{\mu}{T}\right)+\sigma E^{\mu}+\xi\, \omega^{\mu}+\,\xi_B B^{\mu}\,\,\,\,\,\,\,\,\,\,\,\,\,\,\,\,\,\,\,\,\,\,\\ \label{disspart_axial}
	\nu^{\mu}_5&=&  - \sigma_5 T P^{\mu \nu} \partial_{\nu}\left(\frac{\mu}{T}\right)+\sigma_5 E^{\mu}+\xi_5\, \omega^{\mu}+\,\xi_{B5} B^{\mu}
	\end{eqnarray}
	with the vorticity defining as $\omega^{\mu}=\frac{1}{2}\epsilon^{\mu\nu\alpha\beta}u_{\nu}\partial_{\alpha}u_{\beta}$.
	The coefficients $\eta$, $\zeta$, $\sigma$ and $\sigma_5$ are dissipative transport coefficients. In the following we mainly study the non-dissipative fluids. So the only relevant coefficients are  the anomalous transport coefficients   $\xi$ and $\xi_{B}$ corresponding to CVE and CME \cite{Son:2009tf,Landsteiner:2011cp,Neiman:2010zi,Bhattacharya:2011tra,Jian,Sadofyev:2010pr}
	\begin{equation}
	\begin{split}\label{transport coef}
	\xi&=2\mathcal{C}\, 
	\left(\mu \mu_5-\frac{n \mu_5}{3w}\left(3\mu^2+\mu_5^2\right)\right)-2\mathcal{D}\,\frac{n \mu_5}{w}T^2\\
	\xi_5&=\mathcal{C}\, 
	\left(\mu^2+ \mu_5^2-\frac{2 n_5 \mu_5}{3 w}\left(3 \mu^2+\mu_5^2\right)\right)+\mathcal{D}\left(1-\frac{2 n_5 \mu_5}{w}\right)T^2\\
	\xi_B&=\mathcal{C}\, \mu_{5}\left(1-\frac{ n \mu}{w}\right)\\
	\xi_{5B}&=\mathcal{C}\mu\left(1-\frac{ n_5 \mu_5}{w}\right)
	\end{split}
	\end{equation}
	where the corresponding chiral anomaly and gravitational anomaly coefficients are:
	\begin{equation}
	\mathcal{C}=\frac{1}{2 \pi^2},\,\,\,\,\,\,\,\,\,\,\mathcal{D}=\frac{1}{6}.
	\end{equation}
	Hydrodynamic excitations are low energy long wavelength
	excitations around the equilibrium state in fluid. To find their dispersion relations, one  has to firstly choose a set of hydrodynamic variables and then  linearize the equations of motion in terms of their fluctuations around a thermodynamic solution. It is  conventional to consider the microscopic conserved quantities and choose the hydro variables associatively, like  
	\begin{equation}
	\phi_a=(\epsilon, \,\boldsymbol{\pi_i},\, n,\, n_5),\,\,\,\,\,\,\,\,\,\,\,\,a=1,2,...,6
	\end{equation}
	where $\epsilon$, $\boldsymbol{\pi}$ and  $n$ have microscopic definitions given by $T^{00}(x)$, $T^{0i}(x)$ and $J^{0}(x)$ \cite{Kovtun:2012rj,Abbasi:2015nka}. However, we prefer to choose $\phi_a$ as it follows:
	\begin{equation}\label{phia}
	\phi_a=(T, \,\boldsymbol{\pi_i},\, \mu,\, \mu_5),\,\,\,\,\,\,\,\,\,\,\,\,a=1,2,...,6
	\end{equation}
	where $\pi_{i}= w v_i$  and $w=\epsilon+p$ is the enthalpy density. This special choice makes the computations simpler when finding the hydrodynamic modes for a fluid with a general equation of state. 
	However, the QCD equation of state obtained from the lattice calculations shows that, at high enough temperatures,  QCD plasma is thermodynamically conformal. The non-conformality of QCD  becomes serious at and just above the critical temperature $T_c$ \cite{CasalderreySolana:2011us}. So the results found by using the conformality assumptions is more quantitatively reliable when applied to data of LHC than when applied to those of RHIC. In this paper,  we focus on QCD plasma at high enough temperatures with the following equation of state:
	\begin{equation}
	\epsilon=\frac{1}{c_s^2} p,\,\,\,\,\,\,\left(c_s=\frac{1}{\sqrt{3}}\right).
	\end{equation}
	The thermodynamic solution in our system  is given by
\footnote{We use the metric $g_{\mu \nu}=(-1,1,1,1)$ in this paper.}
	\begin{equation}
	\begin{split}
	& u^{\mu}=\left(1,\frac{}{}\boldsymbol{\Omega} \times \boldsymbol{x}\right)\,\,\,\,\,\,\,\,\,\,\,\,\,\,\,\Omega\, r \ll 1,\\
	\,\,& T=Const.,\,\,\mu=Const.,\,\,\mu_5=Const.\\
	&\boldsymbol{B}=Const.
	\end{split}
	\end{equation}
	with $r$ being the distance from the rotation axis. The pressure $p=p(T,\mu, \mu_5)$ satisfies:
	\begin{eqnarray}
	dp&=& s dT+ n d\mu+n_5d \mu_5.
	\end{eqnarray}
	In this paper, we compute six hydrodynamic modes associated with six hydro variables in three different cases. Except in a subsection related to QCD fluid  in quark-gluon-plasma experiments, we always neglect the effect of dissipation in our study. We first consider hydro excitations in the equilibrium of the QCD fluid coupled to an external magnetic field ($\boldsymbol{B}\ne0, \boldsymbol{\Omega}=0$). We then turn off the magnetic field and consider the hydro excitations in the QCD matter rotating with a constant vorticity ($\boldsymbol{B}=0, \boldsymbol{\Omega}\ne0$). Finally in the most general case, the hydro modes are studied in rotating  QCD fluid coupled to an external magnetic field.  
	Let us mention that in the last part of the paper, we take into account the effect of dissipation  in the special case where $\boldsymbol{\Omega}=0$ and $\mu_5=0$.  Let us note that the effect of dissipation has been also considered in 		\cite{Huang:2013iia} to study the induction of axial current in the direction of electric field in thermal QED plasma.    
	
	%
	
	\section{QCD Fluid Coupled to External Magnetic Field}
	\label{section_B}
	In this section, we consider a QCD type fluid coupled to an external magnetic field and compute the  spectrum of its hydrodynamic excitations in detail. After deriving the covariant linearized equations, we divide our computations into two parts. First, we consider pure scalar perturbations and then in another subsection, we take the mixed scalar-vector perturbations under study. To be more complete and clear, we also discuss on the Riemann invariants  of the fluid and show, to each hydro excitation,  which coherent combination of perturbations corresponds. 
	\subsection{Equations of Motion Linearized}
	\label{section linear}
	Let us consider the hydro field defined in (\ref{phia}) is slightly deviated from its thermodynamic value as the following:
	\begin{equation}
	\phi_{a}+\delta \phi_a=\left(T+\delta T, \,\frac{}{}\boldsymbol{0}+\boldsymbol{\pi},\,\mu+\delta \mu,\,\mu_5+\delta \mu_5\right).
	\end{equation}
	To first order in $\delta$ variations, the equations of motion may be covariantly written as
	\begin{equation}\label{linear_EOM}
	M^{\boldsymbol{B}}_{ab}(\boldsymbol{k} , \omega)  \delta \phi_b (\boldsymbol{k} , \omega) = 0
	\end{equation}
	with $M^{\boldsymbol{B}}_{ab}$ given by (see \ref{susceptibility matrix} for thermo coefficients):
	\begin{equation*}\label{MatrixB}
	\resizebox{1.05\hsize}{!}{$
		\begin{bmatrix}
		- i \alpha_1\omega & i k_j & -i \alpha_2 \omega & -i \alpha_3 \omega \\
		i\alpha_1  v_s^2 k^i & -i\omega \delta^i_j -i \frac{\xi}{2\bar{w}} \left(\boldsymbol{B} \cdot \boldsymbol{k} \delta^i_j - B_j k^i \right) - \frac{\bar{n}}{\bar{w}} \epsilon^i \,_{jl}B^l &  i\alpha_2  v_s^2 k^i & i\alpha_3  v_s^2 k^i \\
		-i \beta_1 \omega+\left(\frac{\partial \xi_B}{\partial T}\right)  i \boldsymbol{B} \cdot \boldsymbol{k} & \frac{\bar{n}}{\bar{w}} i k_j - \frac{\xi_B}{\bar{w}} i\omega B_j & -i\beta_2\omega +  \left(\frac{\partial \xi_B}{\partial \mu}\right) i \boldsymbol{B} \cdot \boldsymbol{k} & -i \beta_3 \omega+ \left(\frac{\partial \xi_B}{\partial \mu_5}\right) i \boldsymbol{B} \cdot \boldsymbol{k}\\
		-i \gamma_1 \omega+\left(\frac{\partial \xi_{5 B}}{\partial T}\right)  i \boldsymbol{B} \cdot \boldsymbol{k} & \frac{\bar{n_5}}{\bar{w}} i k_j - \frac{\xi_{5 B}}{\bar{w}} i\omega B_j &  -i \gamma_2 \omega+ \left(\frac{\partial \xi_{5 B}}{\partial \mu}\right) i \boldsymbol{B} \cdot \boldsymbol{k} & - i\gamma_3  \omega+ \left(\frac{\partial \xi_{5 B}}{\partial \mu_5}\right) i \boldsymbol{B} \cdot \boldsymbol{k}
		\end{bmatrix}.
		$}
	\end{equation*}
		The superscript $\boldsymbol{B}$ on $M^{\boldsymbol{ B}}_{ab}$ here refers to this point that in this section  we are studying modes in presence of an external magnetic field. In the next two sections, we change the superscripts to $\boldsymbol{ \Omega}$ and $\boldsymbol{B \Omega}$ respectively.
	As it can be obviously seen above, none of the elements of matrix $M^{\boldsymbol{B}}_{ab}$ vanishes in general. It means that each of the hydrodynamic excitations in this system might be  a coherent excitation of all scalar and vector perturbations. Via computing the Riemann invariants, however, one can exactly determine the type of each propagating modes in the fluid\footnote{Throughout this paper wherever we mention scalar or vector, we mean the representations of $SO(3)$ spatial rotational group.}.
	\subsection{Characteristics and Riemann Invariants}
	Before starting to compute the hydrodynamic modes, let us briefly review the notion of characteristics and the  Riemann invariants in fluid dynamics. As it is well-known, in a fluid whose space of states is $d-$dimensional (in our case $d=6$), there exist in general $d$ characteristics or equivalently $d$ hydrodynamic waves.
	These characteristics describe the different ways through which, a small perturbation  in the state of fluid may propagate in the state-space. To each of the characteristics, one family of integral curves in the state-space is corresponded.  
	Those perturbations that propagate only through curves of one characteristic family correspond to the Riemann invariants \cite{Landau}.  So the Riemann invariant $\mathcal{R}_i$ associated with the hydro mode $\omega_{i}$ satisfies the following equation:
	\begin{equation}\label{RiemannDefinition}
	\left(\partial_t+\frac{}{}\boldsymbol{v}_i \cdot \boldsymbol{\nabla}\right)\mathcal{R}_i=0, \,\,\,\,\,\,\,i=1,\dots, 6
	\end{equation}
	where $\boldsymbol{v}_i$ is the velocity of $i^{th}$ mode, namely $\omega_i$.
	
	In order to determine the Riemann invariants, one assumes that the linear equations of perturbations  may be written as the following:
	\begin{equation}
	\partial_{t} \delta \phi_{a}(\boldsymbol{k},t)+\,D^{\boldsymbol{ B}}_{ab}(\boldsymbol{k})\, \delta \phi_b(\boldsymbol{k},t)=\,0.
	\end{equation}
	Firstly, it is needed to compute the eigenvalues of the matrix $D^{\boldsymbol{B}}$ as the characteristics or equivalently the hydrodynamic modes. To this end one has to find the roots of the determinants of the matrix $M^{\boldsymbol{B}}$, perturbatively order by order, in derivative expansion. The structure of eigenmodes is as the following:
	\begin{equation}\nonumber
	\omega^{\boldsymbol{B}}_i(\boldsymbol{k},\boldsymbol{B})=\omega_i^{\boldsymbol{B}(1)}(\boldsymbol{k},\boldsymbol{B})+\omega_i^{\boldsymbol{B}(2)}(\boldsymbol{k},\boldsymbol{B})+\dots.
	\end{equation}
 In the above equation, $\omega_i^{\boldsymbol{ B}(1)}$ and  $\omega_i^{\boldsymbol{ B}(2)}$,  are the zero and  first order derivative parts of dispersion relation. More explicitly,  if we get $\epsilon_f$ as the parameter which counts the number of derivatives, we would have 
	\begin{eqnarray}\label{rescaling}\nonumber
	\omega_i^{\boldsymbol{B}(1)}(\epsilon_f \boldsymbol{k}, \epsilon_f \boldsymbol{B})&=&\epsilon_f\, \omega_i^{\boldsymbol{B}(1)}(\boldsymbol{k},\boldsymbol{B})\\\label{rescaling2}
	\omega_i^{\boldsymbol{B}(2)}(\epsilon_f \boldsymbol{k}, \epsilon_f \boldsymbol{B})&=&\epsilon_{f}^2\, \omega_i^{\boldsymbol{B}(2)}(\boldsymbol{k},\boldsymbol{B})\\\nonumber
	\cdots& =& \cdots.
	\end{eqnarray}
	
	The next step is to compute the eigenvectors of matrix $D^{\boldsymbol{ B}}$. Then the Riemann invariant associated with  each of these vectors  is  the special  scalar combination of the components of $\delta \phi_a$ which remains invariant along the integral curve generated by that eigenvector in the space-state.  It should be denoted that in Fourier space, $D^{\boldsymbol{ B}}_{ab}=M^{\boldsymbol{ B}}_{ab}+i \omega \delta_{ab}$ where $M^{\boldsymbol{B}}_{ab}$ was defined in (\ref{linear_EOM}).  
	
	\subsubsection{   Eigenvectors}
	In general, a hydrodynamic mode  with dispersion relation $\omega_i=\omega_i(k)$ is characterized as  a plane wave
	\begin{equation}
	\delta\phi_{a \,i}(t,\boldsymbol{x})=\delta \tilde{\phi}_{a \,i} e^{-i \omega_i(k)t+ i \boldsymbol{k.x}}
	\end{equation}
	where the amplitude of the wave, namely $\delta \tilde{\phi}_i$, is referred to as the $i^{th}$ eigenvector of the matrix $M^{\boldsymbol{ B}}$.
	The basis for our six dimensional state-space is made out of $\delta T$, $\delta \boldsymbol{\pi}$, $\delta \mu$ and $\delta \mu_5$. So a general eigenvector of the matrix $M^{\boldsymbol{ B}}$ takes the following form in this basis:
	\begin{equation}
	\delta \tilde{\phi}_{a \,i}=\left(\delta \tilde{\phi}_1, \frac{}{}\delta\tilde{\phi}_{2,3,4}, \,\delta \tilde{\phi}_{5},\, \delta \tilde{\phi}_{6}\right)_i:=\,\left(\delta \tilde{\phi}_T, \frac{}{}\delta\tilde{\phi}_{\boldsymbol{\pi}}, \,\delta \tilde{\phi}_{ \mu},\, \delta \tilde{\phi}_{\mu_5}\right)_i.
	\end{equation}
	
	\subsubsection{Different sectors of the propagation}
	Depending on the type of perturbations carrying by a  mode, one can characterize the eigenvectors  into the scalar, vector or mixed sub sectors. As before we use scalar and vector  terminologically for the representations of the $SO(3)$ group orthogonal to the local velocity of the fluid at each point. So the scalar modes are those that carry the perturbations of the $\delta T$, $\delta \mu$ or $\delta \mu_5$ while  a vector mode carries a combination of the momentum perturbations $\boldsymbol{\pi}$s. It is clear that a mixed scalar-vector mode carries scalar perturbations together with the vector perturbations.
	
	Our computations show that in a general fluid with two axial and vector currents, no pure vector type hydrodynamic mode  propagates. We find that the six hydrodynamic modes of the fluid, obtained from the matrix $M^{\boldsymbol{B}}$ in section (\ref{section linear}), are divided into the following two sets:
	
	\textbf{1:} two scalar modes (\ref{CMWB}).

	It turns out that these two modes vanish at zeroth order, namely in ideal hydrodynamics, and appear from the first order in derivative expansion:
	\begin{equation}
	\omega^{\boldsymbol{ B}}_{1,2}=0+\omega^{\boldsymbol{ B}(2)}_{1,2}+O(\partial^3)
	\end{equation}
	
	\textbf{2:} four mixed scalar-vector modes (\ref{AlfsoundB34}) and (\ref{AlfsoundB56}).

	These four modes are in general non-vanishing in both zero and first orders of derivatives:
	\begin{equation}
	\omega^{\boldsymbol{ B}}_{3,4,5,6}=\omega^{\boldsymbol{ B}(1)}_{3,4,5,6}+\omega_{3,4,5,6}^{\boldsymbol{ B}(2)}+O(\partial^3)
	\end{equation}
	In the following two subsections we give the  dispersion relations and also discuss about the nature of the above 1 and 2 sets separately.
	
	\subsection{Scalar Sector: Chiral Magnetic-Heat Waves}
	Among the six eigenmodes, two modes vanish at ideal (zero) order. The first non-vanishing contribution to their dispersion relation comes from the  first order corrections of constitutive relations. One finds:
	\begin{equation}\label{CMWB}
	\omega^{\boldsymbol{B}}_{1,2}(k)=\,-\frac{\mathcal{A}_1\pm\sqrt{\mathcal{A}_1^2-\,\mathcal{A}_2\mathcal{E} } }{\mathcal{E}}\,\boldsymbol{B}\cdot\boldsymbol{k}
	\end{equation}
	where the $\omega_1^{\boldsymbol{B}}$ and $\omega_2^{\boldsymbol{B}}$ refer to $+$ and $-$ in front of the square root, respectively. We call the velocity of these modes $v_{CMHW1}$ and $v_{CMHW2}$.
	In the above formula, we have defined 
	\begin{equation}
	\mathcal{E}=-\epsilon^{ijk} \alpha_i \beta_j \gamma_k\,\,\,\,\,\,\,\,\,(\epsilon^{123}=1)
	\end{equation}
	with anomaly expressions \footnote{$
		A_{[i}B_{j]}=A_i B_j-A_jB_i$.}
	\begin{eqnarray}
	\mathcal{A}_0&=&\frac{n \mu}{w}\alpha_{[1}\gamma_{2]}+\frac{n_5 \mu_5}{w}\alpha_{[1}\beta_{3]}-\frac{n \mu_5}{w}\alpha_{[1}\gamma_{3]}-\frac{n_5 \mu}{w}\alpha_{[1}\beta_{2]},\\\label{A1}
	\mathcal{A}_1 &=&\frac{\mathcal{C}}{2}\left(\alpha_{[3} \beta_{1]}+\alpha_{[2} \gamma_{1]}+\frac{2\mu\mu_5}{w}\mathcal{E}+\mathcal{A}_0\right),\\\label{A2}
	\mathcal{A}_2&=&\mathcal{C}^2 \alpha_1\left(1-\frac{n \mu+ n_5 \mu_5}{w}\right)+\mathcal{C}^2\frac{\mu\mu_5}{w}\left(\alpha_{[3} \beta_{1]}+\alpha_{[2} \gamma_{1]}+\frac{\mu\mu_5}{w}\mathcal{E}+\mathcal{A}_0\right).
	\end{eqnarray}
	
\begin{figure}
			\centering
			\begin{tabular}{cccc}
\subfloat[][$\frac{\mu_5}{T}=3$ and $\frac{B}{T^2}=0.5$\label{fig:sfig1}]{\includegraphics[width=7 cm]{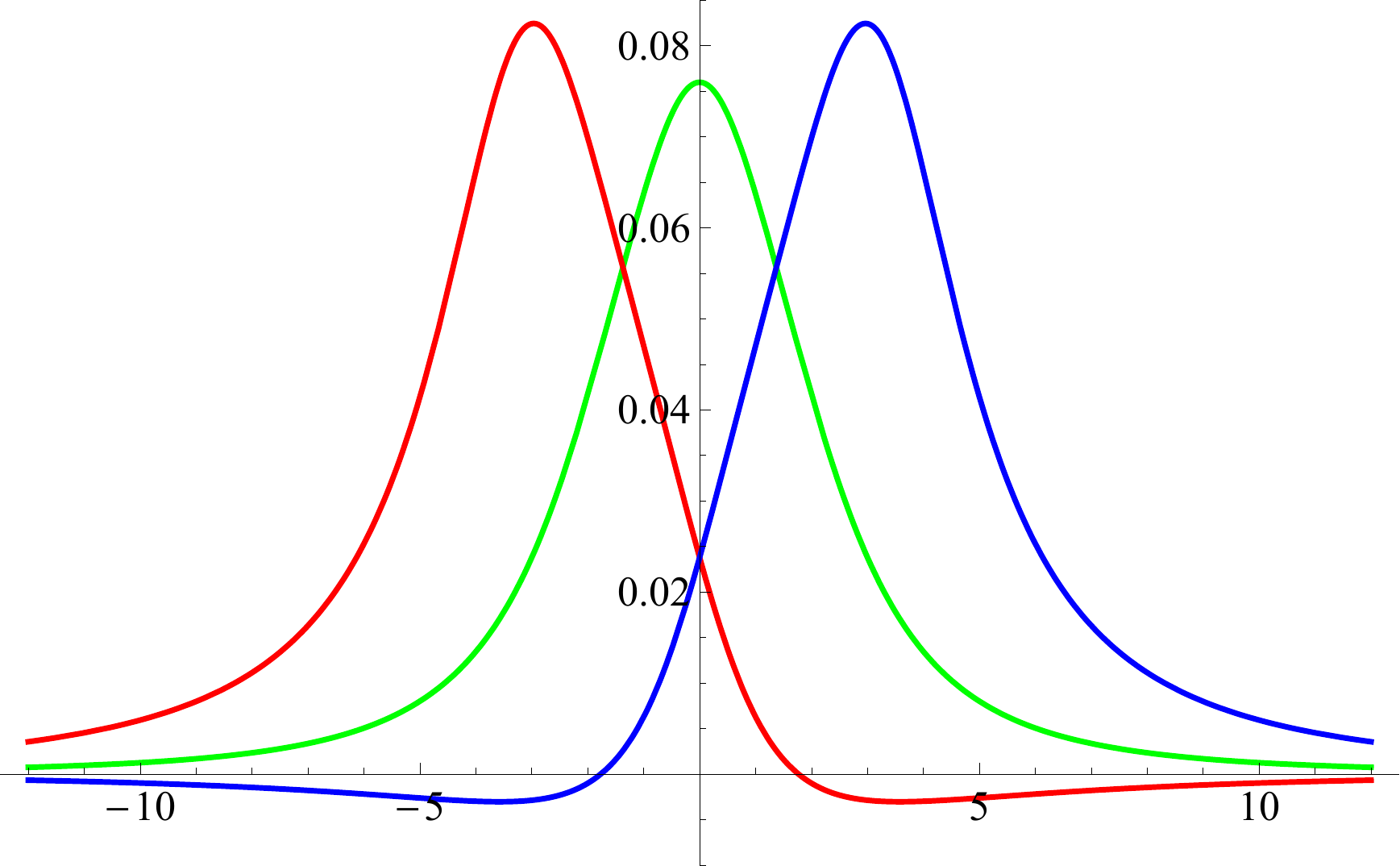}} &	
		\put(-30,0){$\mu/T$}
		\put(-110,130){$v_{CMW}$}
\subfloat[][$\frac{\mu_5}{T}=1$ and $\frac{B}{T^2}=0.5$ \label{fig:sfig2}]{%
  \includegraphics[width=7cm]{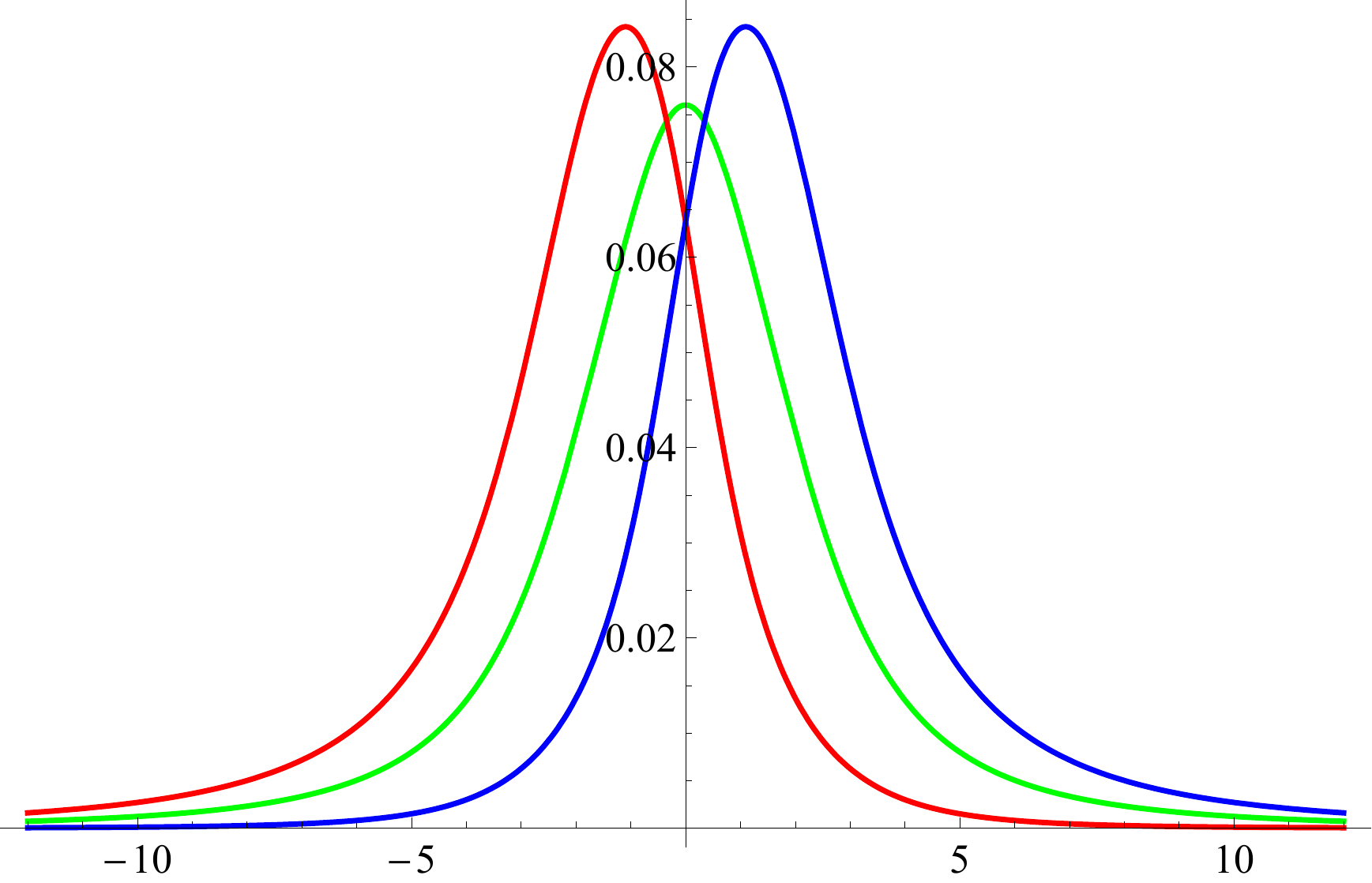}}&			
		\put(0,0){$\mu/T$}
				\put(-110,130){$v_{CMW}$}
		\end{tabular} 
		\caption{The velocities of the non-degenerate CMHWs (red$=v_{CMHW1}$ and blue$=-v_{CMHW2}$) compared to the case $\mu_5=0$ (green). In both panels, we have taken $\boldsymbol{k} \parallel \boldsymbol{B}$. }
		\label{fig1}
	\end{figure}

		It is worth mentioning that depending on the value of $\mu$ and $\mu_5$, the overall sign of each mode in \eqref{CMWB}  might be either positive or negative. The probable minus sign in the dispersion relation means that in order to have positive frequency, the wave has to propagate in the opposite direction of a mode with positive sign. This relative behavior can be clearly seen in Figure \eqref{fig1}. The same argument goes on for other similar situations in this paper.
	
	The  eigenvectors associated with modes \eqref{CMWB} are
	\begin{equation}\label{eigen_vec_B_CMW}
	\delta \tilde{\phi}_{1,2}^{\boldsymbol{ B}}=\,\left(r\,\frac{\alpha_2}{\alpha_1}+s\,\frac{\alpha_3}{\alpha_1},\,\boldsymbol{0},\, -r,\, -s\right),
	\end{equation}
	with $r$ and $s$ being arbitrary parameters. Let us mention that these vectors have been given up to zero order in derivative expansion. The ambiguity in fully specifying these eigenvectors is related to this point that to this order, the modes $\omega^{\boldsymbol{ B}}_{1,2}$ are degenerate, both with the eigenvalue being zero. So  we have the freedom to choose any two arbitrary vectors with the above form as the corresponding eigenvectors. One can find two linearly independent orthogonal eigenvectors $	\delta \tilde{\phi}_{1,2}^{\boldsymbol{ B}}$ as the following.
First we take two vectors from the subspace spanned by \eqref{eigen_vec_B_CMW} by choosing $r=1$, $s=0$ and $r=0$, $s=1$: 	
\begin{eqnarray}
n^{\boldsymbol{B}}_1&=&\left(\frac{\alpha_2}{\alpha_1},\,\,\boldsymbol{0},\,\,-1,\,\,0\right)\\
n^{\boldsymbol{B}}_2&=&\left(\frac{\alpha_3}{\alpha_1},\,\,\boldsymbol{0},\,\,-1,\,\,0\right).
\end{eqnarray}
	Now we project $n^{\boldsymbol{B}}_2$ on the direction of $n^{\boldsymbol{B}}_1$ and then subtract the projection vector from $n^{\boldsymbol{B}}_2$. The resultant vector, $n^{\boldsymbol{B}}_3$, is perpendicular to $n^{\boldsymbol{B}}_1$. So we can get this vector together with $n^{\boldsymbol{B}}_1$ as the two eigenvectors corresponding to CMWVs:
	\begin{eqnarray}\label{eigenvec}
	\delta \tilde{\phi}_{1}^{\boldsymbol{ B}}&=&n^{\boldsymbol{B}}_1=\,\left(\frac{\alpha_2}{\alpha_1},\,\,\boldsymbol{0},\,\,-1,\,\,0\right)\\\label{eigenvec2}
	\delta \tilde{\phi}_{2}^{\boldsymbol{ B}}&=&n^{\boldsymbol{B}}_3=\,\left(\frac{\frac{\alpha_3}{\alpha_1}}{1+\big(\frac{\alpha_2}{\alpha_1}\big)^2},\,\,\boldsymbol{0},\,\,\frac{\frac{\alpha_3 \alpha_2}{\alpha_1^2}}{1+\big(\frac{\alpha_2}{\alpha_1}\big)^2},\,\,-1\right).
	\end{eqnarray}
	That the amplitude of these waves is spanned by $\delta T$, $\delta n$ and $\delta n_5$ in the $3$-dimensional scalar subspace of state-space means that these modes are scalar-type.
	Let us recall that by using the standard thermodynamic transformations (see Appendix \ref{Appen thermo})
	one can alternatively express the eigenvectors (\ref{eigen_vec_B_CMW}) in terms of another set of fluctuations, e.g.  $\delta \epsilon$, $\delta n$ and $\delta n_5$.  
	So the modes $\omega^{\boldsymbol{ B}}_{1,2}$  are in fact the coherent perturbations of energy, vector and axial charge currents.
	
	Let us remind that while the well-known CMWs \cite{Kharzeev:2010gd} carry exclusively the perturbations of axial and vector currents, the waves found here carry the energy (temperature) perturbations as well. 
	For this reason,  we refer to them as the Chiral Magnetic-Heat wave (CMHW).  In order to make clear the feature of CMHWs, let us go back and consider the eigenvectors (\ref{eigen_vec_B_CMW}).  The thermodynamic coefficients $\alpha_2=\partial{\epsilon}/\partial{\mu}$ and
	$\alpha_3=\partial{\epsilon}/\partial{\mu_5}$ are non-vanishing at finite vector and axial charge densities, so one would expect only at $n=n_5=0$  \cite{Kharzeev:2010gd}, the temperature perturbation is not carried by the CMHWs (see equation (\ref{eigen_vec_B_CMW})). In the latter limit, the CMHW changes to CMW. Additionally, while both the  left- and right-moving CMWs in \cite{Kharzeev:2010gd} are identified with one velocity,  the velocity of two CMHWs is not the same; one of them in general moves faster than the other. It is in fact the manifestation of the energy transport by the  CMHWs. 
	
	Using the exact form of the equation of state, the above discussion can be understood more quantitatively.  In a conformal fluid of non-interacting fermions with both vector and axial charges, we have 
	\begin{equation}\label{EoS}
	\epsilon=3 p=\frac{7 \pi^2}{60} T^4+\frac{1}{2}(\mu^2+\mu_5^2)T^2+\frac{1}{4\pi^2}(\mu^4+6 \mu^2 \mu_5^2+\mu_5^4).
	\end{equation} 
	In the following, we study CMHWs found above in a fluid with the above equation of state.
	In Figure \eqref{fig1}, we have plotted the dependence of the velocities of the fast and slow CMHWs on $\frac{\mu}{T}$, in two cases, each of which corresponded to a special value of the $\frac{\mu_5}{T}$ in equilibrium.\footnote{We work in a system of units where $c=1$.} In each case, we have also depicted the changes of CMHW at $\mu_5=0$ with a green curve. Note that according to \eqref{CMWB}, CMHWs are not dispersive and this means that the curves presented in this figure are unique for CMHWs with each arbitrary wavelength in the hydrodynamic regime.
	
	As it is observed in the Figure, for large values of $\frac{\mu_5}{T}$, the fast and slow waves may propagate either in opposite or in the same directions (panel a); it depends on the value of $
	\frac{\mu}{T}$ in a fixed $\frac{B}{T^2}$. In the smaller values of $\frac{\mu_5}{T}$, however, two CMHWs always propagate opposite to each other (panel b). It should be noted that in both panels, the difference between velocities of the fast and slow  CMHWs,  is due to presence of a finite axial charge density in the fluid; actually at $\mu_5=0$, the coefficient $\mathcal{A}_1$
	in  (\ref{CMWB}) vanishes and the velocities of both waves become the same.

	Another point with figure 1 is that each of the fast and slow waves reaches to its maximum velocity when $\mu=\mu_5$ or $\mu=-\mu_5$ \footnote{it  can be simply obtained by solving $\partial_{\nu} v_{CMHW}=0$ with $\nu=\mu/T$.}. Consequently, when $\mu_5=0$, the velocities of two waves become degenerate with a maximum at $\mu=\mu_5=0$ (green curve). 
	
	As the last point regarding the figure 1, we compare two special limits with each other. First suppose $n_5=0$ while $n\ne0$; in this case, as can be clearly seen by green curves in figure, a degenerate CMHW does exist. In the opposite limit when $n_5\ne0$ and $n=0$, again two degenerate CMHW propagate corresponding with the common $v$-intercept of blue and red curves in  the figure.

	Before ending this subsection, let us emphasize that the novelty of our results is not limited to the $\mu_{5}\ne0$ case. Even at $\mu_{5}=0$, our results are novel since 
	we have considered the fluctuations of energy-momentum as well as the charge fluctuations. To make this point clear, let us consider equation \eqref{CMWB}. At $\mu_{5}=0$ this equation simplifies to
	\begin{equation}\label{CMWmu0}
	\omega^{\boldsymbol{B}}_{1,2}=\pm \frac{B k}{2\pi^2 \chi} \frac{1-\frac{\mu n}{w}}{\sqrt{1-\frac{\mu n }{w}-\frac{n}{\chi w}\left(\frac{n}{c_s^2}-\chi \mu\right)}}
	\end{equation}
	with $\chi=\partial n/\partial\mu$. This result differs clearly from the CMW 
	\begin{equation}\label{kharz}
	\omega=\pm \frac{B k}{2\pi^2 \chi} 
	\end{equation}
	obtaining in \cite{Kharzeev:2011ds} by turned off  energy-momentum fluctuations \footnote{Some part of this difference might be related to difference between choices of frame in our paper and \cite{Kharzeev:2011ds}. We will shed light on the issue in \cite{Abbasi1}.}.

	\subsection{Mixed Scalar-Vector Sector}
	In addition to two scalar modes given in previous subsection, matrix $M^{\boldsymbol{ B}}_{ab}$ has another four perturbative roots corresponding to four different hydro modes. In contrast to scalar sector modes, the four new modes are present even in ideal (zero order) hydrodynamics. Up to first order in derivative correction of constitutive relations, or equivalently up to second order in derivative expansion of dispersion relations, we obtain $\omega^{\boldsymbol{ B}}_{3,4}=\,\omega^{\boldsymbol{B}(1)}_{3,4}+\,\omega^{\boldsymbol{B}(2)}_{3,4}$ with
	\begin{equation}\label{AlfsoundB34}
	\begin{split}
	\omega^{\boldsymbol{B}(1)}_{3,4}=&\pm\frac{1}{\sqrt{2}}\sqrt{c_s^2  k^2+\Omega_{L}^2+\sqrt{\big(c_s^2 k^2+\Omega_{L}^2\big)^2-4 c_s^2 k^2\, \Omega_{L}^2\cos^2\theta}}\\
	\omega^{\boldsymbol{B}(2)}_{3,4}=&-\frac{\Omega_{L}^2\left( \big(\omega_{3,4}^{\boldsymbol{B}(1)}\big)^2-\frac{}{}c_s^2 k^2\cos^2\theta\right)\,\cos \theta }{(c_s^2 k^2+\Omega_{L}^2)\big(\omega_{3,4}^{\boldsymbol{ B}(1)}\big)^2-2 c_s^2 k^2\,\Omega_{L}^2 \cos^2\theta}\,\left(\frac{\xi\frac{}{} B}{2 w}\right)k
	\end{split}
	\end{equation}
	and $\omega^{\boldsymbol{B}}_{5,6}=\,\omega^{\boldsymbol{B}(1)}_{5,6}+\,\omega^{\boldsymbol{B}(2)}_{5,6}$ with
	\begin{equation}\label{AlfsoundB56}
	\begin{split}
	\omega^{\boldsymbol{B}(1)}_{5,6}=&\pm\frac{1}{\sqrt{2}}\sqrt{c_s^2  k^2+\Omega_{L}^2-\sqrt{\big(c_s^2 k^2+\Omega_{L}^2\big)^2-4 c_s^2 k^2\, \Omega_{L}^2\cos^2\theta}}\\
	\omega^{\boldsymbol{B}(2)}_{5,6}=&-\frac{\Omega_{L}^2\left( \big(\omega_{5,6}^{\boldsymbol{ B}(1)}\big)^2-\frac{}{}c_s^2 k^2\cos^2\theta\right)\,\cos \theta }{(c_s^2 k^2+\Omega_{L}^2)\big(\omega_{5,6}^{\boldsymbol{ B}(1)}\big)^2-2 c_s^2 k^2\,\Omega_{L}^2 \cos^2\theta}\,\left(\frac{\xi\frac{}{} B}{2 w}\right)k
	\end{split}
	\end{equation}
	where $\cos \theta=\boldsymbol{\hat{B}}\cdot\boldsymbol{\hat{k}}$. In the equations given above, $\Omega_{L}$ is the Larmor frequency as being
	\begin{equation}\label{Larmor}
	\Omega_{L}=\,\frac{n B}{w}.
	\end{equation}
	Considering \eqref{rescaling},  the outer square root in $\omega^{(1)}_{3,4,5,6}$ turns out to be  of order $O(\partial)$. In the case of $\omega^{(2)}_{3,4,5,6}$ however, more clarification is needed to understand why it is of order $O(\partial^2)$. Note that under rescaling $k\rightarrow \epsilon_f k$ and $B\rightarrow \epsilon_f B$, the fraction part in these relations behaves as a zero order object (fraction$\rightarrow \epsilon_f^0$ fraction). So the same as for $\frac{\xi B}{w}k$, the derivative order of $\omega^{(2)}_{3,4,5,6}$ is $O(\partial^2)$.
	
	The eigenvectors corresponding to the above four modes are as the following 
	\begin{equation}\label{eigenvec3456B}
	\begin{split}
	\delta \tilde{\phi}_i^{\boldsymbol{B}}&=\left(\delta T,\,\delta \boldsymbol{\pi},\frac{}{}\delta \mu,\, \delta \mu_5\right)\\
	&=\left(1,-\frac{w \mathcal{E}\,\omega_i^{(1)}\,\boldsymbol{k}}{C_2 \boldsymbol{k}^2}+\frac{n \mathcal{E}\left(i (\omega_i^{\boldsymbol{B}(1)})^{2}\,(\boldsymbol{B}\times\boldsymbol{k})+\frac{n}{w}\omega_i^{\boldsymbol{B}(1)}(\boldsymbol{B}\cdot\boldsymbol{ k})\boldsymbol{B}-\frac{n}{w}\omega_i^{\boldsymbol{B}(1)}(\boldsymbol{B} \cdot \boldsymbol{\hat{k}})^2\boldsymbol{k}\right)}{C_2\big((\omega_i^{\boldsymbol{B}(1)})^2 \boldsymbol{k}^2 -n^2/w^2 \,(\boldsymbol{B} \cdot \boldsymbol{k})^2\big)},\frac{C_1}{C_2},\frac{C_3}{C_2}\right)
	\end{split}
	\end{equation}
	with
	\begin{eqnarray}\label{C_1}
	C_1&=&n\, \alpha_{[1}\gamma_{3]} -n_5\, \alpha_{[1}\beta_{3]}-w\, \beta_{[1}\gamma_{3]} \\\label{C_2}
	C_2&=& n\, \alpha_{[3}\gamma_{2]}-n_5\, \alpha_{[3}\beta_{2]}-w\, \beta_{[3}\gamma_{2]} \\\label{C_3}
	C_3&=&  n\, \alpha_{[2}\gamma_{1]}-n_5\, \alpha_{[2}\beta_{1]}-w\, \beta_{[2}\gamma_{1]}.
	\end{eqnarray}
 Let us denote that the above eigenvectors have generally non-vanishing scalar and vector components in the  state-space. For this reason, we refer to the current sector as the scalar-vector sector. 
	
	At zero order in derivative expansion, each of these modes is a mixture of the ordinary sound with Larmor frequency, reminiscent of the magnetosonic waves in the ideal magnetohydrodynamics \cite{Schnack}.
	At first order in derivative expansion, these four mixed modes get corrections proportional to the magnetic field. The situation is actually analogous to what appears in the dispersion relation of Chiral-Alfv\'en-Waves (CAW) in a chiral fluid of single right-handed fermions \cite{Yamamoto:2015ria,Abbasi:2015saa}. 
	As a result, one may refer to the scalar-vector sector modes as the mixed Sound-Alfv\'en waves.  In the following, it becomes more clear why this terminology is used.

	In the special case of propagation in the direction of magnatic field, $\boldsymbol{B}\parallel \boldsymbol{k}$, the above scalar-vector modes become distinguishable with the following velocities:
	\begin{eqnarray}
        \omega^{\boldsymbol{B}}_{3,4}=\pm c_s k\,\,\,\,\,\,\,\,\,\rightarrow\,\,\,\,\,\,\,\,\,\,v^{\boldsymbol{B}}_{3,4}&=&\pm c_s \\
	\omega^{\boldsymbol{B}}_{5,6}=\pm\frac{n B}{w}-\frac{\xi}{2 w}\,B.\label{} k\,\,\,\,\,\,\,\,\,\,\rightarrow\,\,\,\,\,\,\,\,\,\,v^{\boldsymbol{B}}_{5,6}&=&-\frac{\xi}{2 w}\,B.\label{vcaw}
	\end{eqnarray}
	Clearly the modes $5$ and $6$ are two gapped chiral waves propagating parallel with the magnetic filed. These are the counterpart of CAWs in a chiral fluid with single chirality, recently found in \cite{Yamamoto:2015ria,Abbasi:2015saa}. The terminology, choosing by reference \cite{Yamamoto:2015ria}, might seem a little misleading; there are some differences between CAWs and standard Alfv\'en waves in magnetohydrodynamics (MHD). First, the  Alfv\'en waves in MHD are gappless while in Chiral fluids they are gapped. Second and more important, 
	it is the dynamics of Maxwell fields which leads the Alfv\'en waves propagate in MHD while in our case,  a non-dynamical magnetic field is able to couple to the local fluctuations of vorticity in the chiral fluid and excites a collective motion parallel to itself,  referred to as the chiral Alfv\'en wave in 	\cite{Yamamoto:2015ria}. Despite knowing these differences, since the waves $5$ and $6$ propagate parallel to the magnetic field we follow
	\cite{Yamamoto:2015ria} and call them the chiral Alfv\'en waves. 
	
	In another limit, when $\boldsymbol{B}\perp \boldsymbol{k}$, we have only two gapped plasmon modes \footnote{We would like to thank to referee for pointing the true name of these modes to us.}:
	\begin{eqnarray}\label{magnetosonic}
	\omega^{\boldsymbol{B}}_{3,4}&=&\pm\,\sqrt{c_s^2 k^2 +\Omega_L^2}\,\,\,\,\rightarrow\,\,\,v^{\boldsymbol{B}}_{3,4}=\pm \frac{c_s }{\sqrt{1+\frac{\Omega_L^2}{c_s^2 k^2}}}\\
	\omega^{\boldsymbol{B}}_{5,6}&=&0.
	\end{eqnarray}
	As one naturally expects,  analogous to the case of a chiral fluid with single chirality \cite{Abbasi:2015saa}, the anomaly effects can not be detected in the directions transverse to the magnetic field here. The only modes propagating in transverse directions are the magnetosonic waves.
	By magnetosonic wave here, however, we do not mean exactly the familiar magnetosonic waves in the ideal magnetohydrodynamics. As it is well-known in magnetohydrodynamics, the pressure perturbations produced by Maxwell dynamics intensify  the fluid pressure perturbations, resulting in an excess in the velocity of sound. While in the latter case, the pressure perturbations are intensified due to the compression and rarefaction of the magnetic field lines, in our case however, a constant magnetic field exerts opposite external Lorentz forces on momentum perturbations and decreases the hydrodynamical pressure.

	In summary, we observe that the modes in the scalar-vector sector are in general the coherent perturbation of all six hydro fields. They are mixed-sound-Alfv\'en waves.

	Before ending this section, let us separate the new results of the paper in this part from their well-known counterpart in the literature. To our knowledge, the hydrodynamic excitations of a chiral fluid with both vector and axial currents was studied only in the absence of momentum perturbations before. In other words, the hydro excitations had been computed only for a "Forced" QCD fluid before.  
	None of the six modes \eqref{CMWB},  \eqref{AlfsoundB34} and \eqref{AlfsoundB56}  were  found in previous studies. In the case of  CMHWs \eqref{CMWB}, even at $\mu_5=0$, our result,  namely \eqref{CMWmu0}, was not well-known before. Only at $\mu=\mu_5=0$ in which the temperature perturbations decouple from that of vector and axial currents, the result, namely \eqref{kharz}, exists in the literature.  The latter is nothing but the well-known chiral magnetic wave.  In the case of mixed vector modes \eqref{AlfsoundB34} and \eqref{AlfsoundB56}, the novelty of our results is twofold; first that our results are covariant by this mean that we have found the dispersion relation for propagation in every arbitrary direction with respect to the external magnetic field.  Second, even in $\boldsymbol{B}\parallel \boldsymbol{k}$, the gapped CAW \eqref{vcaw} found in the current paper was not found before, although in the case of single chirality fluid such mode had been found firstly in \cite{Abbasi:2015saa} and afterward in  \cite{Kalaydzhyan:2016dyr} \footnote{Note that the idea of studying the momentum perturbations was firstly in \cite{Yamamoto:2015ria} and then authors of \cite{Abbasi:2015saa} took into account the energy perturbations as well.}.
Furthermore, nowhere in the literature we have seen the eigenvectors \eqref{eigen_vec_B_CMW}  and \eqref{eigenvec3456B} associated with hydro modes 	in a QCD type fluid.

	\section{Rotating QCD Fluid}
	\label{section_Omega}
	In this part, we consider a QCD fluid,  rotating with constant vorticity $\boldsymbol{\Omega}$, in the absence of electromagnetic fields, with the four velocity
	\begin{equation}
	u^{\mu}=\gamma\,\left(1,\frac{}{}\boldsymbol{\Omega} \times \boldsymbol{x}\right).
	\end{equation}
	In what follows, we consider the regime $\Omega\,r \ll1$, where $r$ is the distance from the axis of the rotation. In this regime the Lorentz factor may be expanded as  $\gamma=1+O\left(\left(\Omega\frac{}{}r\right)^2\right)$, so the vorticity  computed in equilibrium up to first order in $O(\Omega r)$ becomes
	\begin{equation}
	\omega^{\mu}=\,\left(0,\frac{}{}\boldsymbol{\Omega}\right).
	\end{equation}
	
	\subsection{Equations of Motion Linearized}
	Let us take the small deviation of hydrodynamic fields (\ref{phia}) away from their equilibrium values as the following
	\begin{equation}
	\delta \phi_a=\left(\delta T, \frac{}{}\boldsymbol{\pi},\,\delta \mu,\,\delta \mu_5\right).
	\end{equation}
	In order to linearize the equations of motion, we have to expand the equations (\ref{EoM}) around the equilibrium state:
	\begin{equation}
	\begin{split}
	&u^{\mu}=\left(1,\frac{}{}\boldsymbol{\Omega} \times \boldsymbol{x}\right)\,\,\,\,\,\,\,\,\,\,\,\,\,\,\,\Omega\, r \ll 1,\\
	\,\,& T=Const.,\,\,\mu=Const.,\,\,\mu_5=Const.
	\end{split}
	\end{equation}
	and keep the terms up to first order in $\delta \phi_a$ fields. Among all terms, there is a delicate point regarding the expansion of vorticity terms of (\ref{disspart_vector}) and (\ref{disspart_axial}) around equilibrium which deserves to be explained in detail. Consider the velocity of fluid is perturbed by $\delta u^{\mu}=\left(\delta u^{0},\frac{}{}\delta\boldsymbol{u}\right)$
	as
	\begin{equation}
	u^{\mu}+\delta u^{\mu}=\left(1+\delta u^0,\frac{}{}\boldsymbol{\Omega} \times \boldsymbol{x}+\delta \boldsymbol{u}\right).
	\end{equation}
	Demanding the above velocity to satisfy the relativistic normalization $u^{\mu}u_{\mu}=-1$, the zero component of the perturbation is immediately fixed
	\begin{equation}
	\delta u^{0}=\left(\frac{}{}\boldsymbol{\Omega} \times\boldsymbol{x}\right)\cdot\,\delta \boldsymbol{u}.
	\end{equation}
	So to first order in perturbations, the vorticity takes the following form 
	\begin{equation}
	\omega^{\mu}+\delta \omega^{\mu}=\,\left(\frac{\xi}{w}\,\big(\delta \boldsymbol{\pi }\cdot \boldsymbol{ \Omega}\big)\,,\,\frac{}{}\boldsymbol{\Omega}+\frac{1}{2 w}\, \boldsymbol{\nabla }\times\delta \boldsymbol{\pi}\right),\,\,\,\,\,\,\,\,\,\,\,\, \Omega\, r \ll 1.
	\end{equation}
	Using the above expression,  the linearized equations of motion may be covariantly written as
	\begin{equation}\label{eq:lin_mat}
	M^{ \boldsymbol{ \Omega}}_{ab}(\boldsymbol{k} , \omega)  \delta \phi_b (\boldsymbol{k} , \omega) = 0,
	\end{equation}
	with $M^{ \boldsymbol{ \Omega}}_{ab}$ given by:
	\begin{equation*}
		\begin{bmatrix}
		- i \alpha_1\omega & i k_j & -i \alpha_2 \omega & -i \alpha_3 \omega \\
		i\alpha_1  v_s^2 k^i & -i\omega \delta^i_j -  \epsilon^i \,_{jl}\Omega^l &  i\alpha_2  v_s^2 k^i & i\alpha_3  v_s^2 k^i \\
		-i \beta_1 \omega+\left(\frac{\partial \xi}{\partial T}\right)  i \boldsymbol{\Omega} \cdot \boldsymbol{k} & \frac{\bar{n}}{\bar{w}} i k_j - \frac{2\xi}{\bar{w}} i\omega \Omega_j & -i\beta_2\omega +  \left(\frac{\partial \xi}{\partial \mu}\right) i \boldsymbol{\Omega} \cdot \boldsymbol{k} & -i \beta_3 \omega+ \left(\frac{\partial \xi}{\partial \mu_5}\right) i \boldsymbol{\Omega} \cdot \boldsymbol{k}\\
		-i \gamma_1 \omega+\left(\frac{\partial \xi_{5 }}{\partial T}\right)  i \boldsymbol{\Omega} \cdot \boldsymbol{k} & \frac{\bar{n_5}}{\bar{w}} i k_j - \frac{2\xi_{5}}{\bar{w}} i\omega \Omega_j &  -i \gamma_2 \omega+ \left(\frac{\partial \xi_{5}}{\partial \mu}\right) i \boldsymbol{\Omega} \cdot \boldsymbol{k} & - i\gamma_3  \omega+ \left(\frac{\partial \xi_{5 }}{\partial \mu_5}\right) i \boldsymbol{\Omega} \cdot \boldsymbol{k}
		\end{bmatrix}.
	\end{equation*}
	At this moment, since all the components of the matrix $M_{ab}^{ \boldsymbol{ \Omega}}$ are non-vanishing,  one may think that each of the characteristics of the fluid is a coherent perturbation of all six scalar and vector hydro variables. We will show in the following that in fact, two of the characteristics are scalar type while the other four are the mixed scalar-vector perturbations.  
	\subsection{Hydro Modes}
	Computing the eigenvalues of matrix $M^{ \boldsymbol{ \Omega}}_{ab}+\,\omega \boldsymbol{1}_{ab}$, or equivalently the roots of $\det M^{ \boldsymbol{ \Omega}}_{ab}=0$, we find six independent hydrodynamic modes of the fluid.   Our computations show that in the rotating fluid, two sets of hydrodynamic modes are present:
	
	\textbf{1:} two scalar modes (\ref{CVW-omega}).

	It turns out that these two modes vanish at zeroth order, namely in ideal hydrodynamics, and just appear from the first order in derivative expansion:
	\begin{equation}
	\omega^{\boldsymbol{\Omega}}_{1,2}=0+\omega_{1,2}^{\boldsymbol{\Omega}(2)}+O(\partial^3)
	\end{equation}
	
	\textbf{2:} four scalar-vector modes (\ref{Rotsound34}) and (\ref{Rotsound56}).

	In contrast to modes in the scalar sector, these four modes are vanishing at first order, contributing at zero order:
	\begin{equation}
	\omega^{\boldsymbol{\Omega}}_{3,4,5,6}=\omega^{\boldsymbol{\Omega}(1)}_{3,4,5,6}+0+O(\partial^3)
	\end{equation}
	In the following two subsections we give the  dispersion relations and also discuss about the nature of the above 1 and 2 sets separately.
	
	\subsubsection{Scalar Sector: Chiral Vortical Heat Waves}
	Among the six eigenmodes, two modes vanish at ideal (zero) order. The first non-vanishing contribution to their dispersion relation comes from the first order corrections of the constitutive relations. One finds:
	\begin{equation}\label{CVW-omega}
	\omega^{\boldsymbol{\Omega}}_{1,2}(k)=\,-\frac{\mathcal{A}_3\pm\sqrt{\mathcal{A}_3^2-
			\mathcal{E}\mathcal{A}_4}}{\mathcal{E}}\;\;\boldsymbol{\Omega}  \cdot\boldsymbol{k}+O(k^3)
	\end{equation}
	where the $\omega_1^{\boldsymbol{\Omega}}$ and $\omega_2^{\boldsymbol{\Omega}}$ refer to $+$ and $-$ in front of the square root, respectively. We also call the velocity of these modes as $v_{CVHW1}$ and $v_{CVHW2}$ respectively.
	The coefficients $\mathcal{A}_3$ and $\mathcal{A}_4$ may be written as polynomials of anomaly coefficients:
	\begin{equation}\label{A3A4}
	\begin{split}
	\mathcal{A}_3&=\mathcal{D}x_1+\,\mathcal{C}x_2\\
	\mathcal{A}_4&=\mathcal{C}^2 y_1+\,\mathcal{D}^2 y_2+\,\mathcal{C D} y_3
	\end{split}
	\end{equation} 
	with $x_i$s  given in Appendix \ref{x} and $y_{i}$s given in Appendix \ref{y}.
	
	The corresponding eigenvectors are 
	\begin{equation}\label{eigen_vec_omega_CVW}
	\delta \tilde{\phi}_{1,2}^{\boldsymbol{\Omega}}=\,\left(r\,\frac{\alpha_2}{\alpha_1}+s\,\frac{\alpha_3}{\alpha_1},\,\boldsymbol{0},\, -r,\, -s\right),\,\,\,\,\,\,\,\,i=1,2
	\end{equation}
	with $r$ and $s$ the arbitrary parameters.  
	Note that we have the freedom to choose any two arbitrary vectors with the above form as the eigenvectors. Analogous to  \eqref{eigenvec} and \eqref{eigenvec2}, we can find two linearly independent orthogonal eigenvectors as the following: 
	\begin{eqnarray}\label{eigenvecvomega}
	\delta \tilde{\phi}_{1}^{\boldsymbol{ \Omega}}&=&\left(\frac{\alpha_2}{\alpha_1},\,\,\boldsymbol{0},\,\,-1,\,\,0\right)\\
	\delta \tilde{\phi}_{2}^{\boldsymbol{ \Omega}}&=&\,\left(\frac{\frac{\alpha_3}{\alpha_1}}{1+\big(\frac{\alpha_2}{\alpha_1}\big)^2},\,\,\boldsymbol{0},\,\,\frac{\frac{\alpha_3 \alpha_2}{\alpha_1^2}}{1+\big(\frac{\alpha_2}{\alpha_1}\big)^2},\,\,-1\right).
	\end{eqnarray}
	Since these two modes carry the perturbations of temperature together with the vector and axial chemical potentials, we call them  Chiral Vortical Heat Waves (CVHW)\footnote{These modes differ from chiral vortical waves found in \cite{Jiang:2015cva}. In the latter reference, the scalar perturbations in a non-chiral limit ($\mu_5=0$) have been found when the temperature kept fixed. }. Generally, one of the CVHWs moves faster than the other. Only in the special limit where the fluid is non-chiral,  namely when $\mu_5=0$ and consequently $\mathcal{A}_3=0$, the velocities of two CVHWs become the same, while definitely propagating in opposite directions \cite{Jiang:2015cva}. 
	\begin{figure}
		\centering
		\includegraphics[width=10cm]{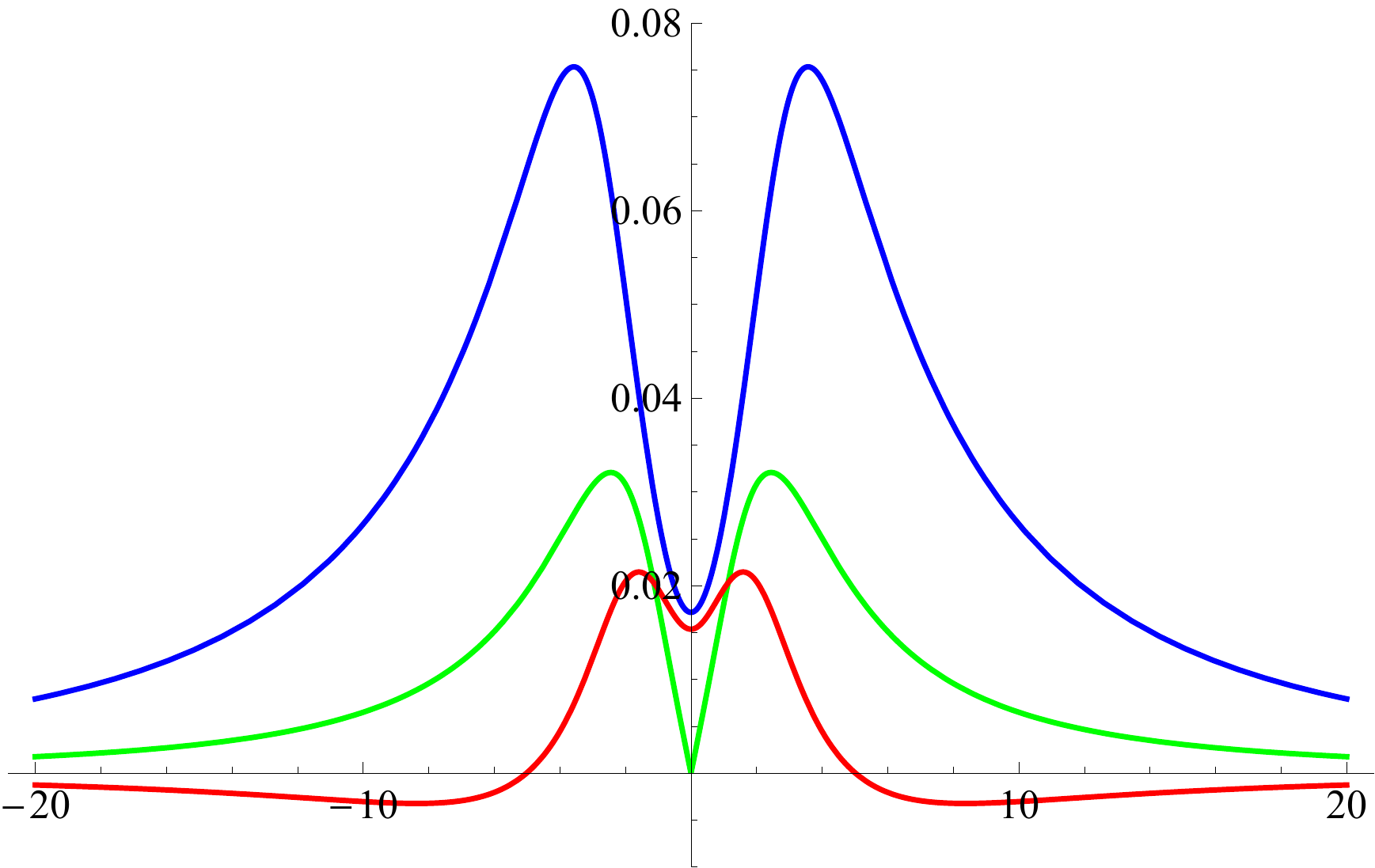}
		\put(0,-2){$\mu/T$}
		\put(-150,190){$v_{CVW}$}
		\caption{The velocities of the non-degenerate CVHWs (red$=v_{CVHW1}$ and blue$=-v_{CVHW2}$) for $\frac{\mu_5}{T}=2$ and $\frac{\Omega}{T}=0.4$ together with the case $\mu_5=0$ (green). We have taken $\boldsymbol{k} \parallel \boldsymbol{\Omega}$. }
		\label{fig2}
	\end{figure}

	Using the equation of state given in \eqref{EoS},
	in Figure \eqref{fig2} we have plotted the dependence of the velocities of the fast and slow CVHWs on $\frac{\mu}{T}$ for a special $\frac{\mu_5}{T}$ and $\frac{\Omega}{T}$ in equilibrium. We have also depicted the changes of CVHW at $\mu_5=0$ with a green curve.
	This plot clearly shows that fast and slow CVHWs do not necessarily propagate in the same direction. As mentioned above, when $\mu_5=0$, the velocity of these two waves become equal to each other, independent of the value of $\frac{\mu}{T}$.
	
	It is worth mentioning that the  nature of CVHWs is different from that of CMHWs in Figure \eqref{fig1}. Interestingly, while CMHWs can propagate in fluid even at $\mu=\mu_5=0$, CVHW can not do so. In addition, the velocity of two CMHWs become degenerate when either $\mu=0$ or $\mu_5=0$. In contrary, CVHWs have the same velocities only when $\mu_5=0$. These observations simply reject this claim that the results in rotating fluid are similar to those in a fluid coupled to magnetic field.  
	This difference is not limited to the scalar sector. In the next subsection, we will see that  the scalar-vector modes in rotating chiral fluid have remarkable differences with those of a non-rotating chiral fluid in a magnetic field. 
	
	\subsubsection{Scalar-Vector Sector}
	In addition to the two scalar modes computed in previous subsection, there are another four modes as the following
	\begin{equation}\label{Rotsound34}
	\omega^{\boldsymbol{\Omega}}_{3,4}=\,\pm\frac{1}{\sqrt{2}}\sqrt{c_s^2  k^2+\Omega^2+\sqrt{\big(c_s^2 k^2+\Omega^2\big)^2-4 c_s^2 k^2\, \Omega^2\cos^2\theta}}+O(k^3)
	\end{equation}
	\begin{equation}\label{Rotsound56}
	\omega^{\boldsymbol{\Omega}}_{5,6}=\,\pm\frac{1}{\sqrt{2}}\sqrt{c_s^2  k^2+\Omega^2-\sqrt{\big(c_s^2 k^2+\Omega^2\big)^2-4 c_s^2 k^2\, \Omega^2\cos^2\theta}}+O(k^3).
	\end{equation}
	with $\cos \theta =\boldsymbol{\hat{\Omega}} \cdot\boldsymbol{ \hat{ k}}$.
	
	Considering \eqref{rescaling},  the square root in all these four modes turns out to be  of order $O(\partial)$ and no second order correction contributes to dispersion of these modes.
	Computing the eigenvectors of the matrix $M^{\Omega}_{ab}$, we find for $i=3,4,5,6$:
	\begin{equation}\label{eigenvec3456Omega}
	\begin{split}
	\delta \tilde{\phi}_i^{\boldsymbol{ \Omega}}&=\left(\delta T,\,\delta \boldsymbol{\pi},\frac{}{}\delta \mu,\, \delta \mu_5\right)\\
	&=\left(1,-\frac{w \mathcal{E}\,\omega^{\boldsymbol{ \Omega}}_i\,\boldsymbol{k}}{C_2 \boldsymbol{k}^2}+\frac{w \mathcal{E}\left(i (\omega_i^{\boldsymbol{\Omega}})^2\,(\boldsymbol{\Omega}\times\boldsymbol{k})+\omega_i^{\boldsymbol{ \Omega}}(\boldsymbol{\Omega} \cdot \boldsymbol{k})\boldsymbol{\Omega}-\omega_i^{\boldsymbol{ \Omega}}(\boldsymbol{\Omega} \cdot \boldsymbol{\hat{k}})^2\boldsymbol{k}\right)}{C_2\big((\omega_i^{\boldsymbol{\Omega}})^2 \boldsymbol{k}^2 - \,(\boldsymbol{\Omega} \cdot \boldsymbol{k})^2\big)},\frac{C_1}{C_2},\frac{C_3}{C_2}\right)
	\end{split}
	\end{equation}
	with $C_1$, $C_2$ and $C_3$ given in (\ref{C_1}), (\ref{C_2}) and (\ref{C_3}).

	In the special case of propagation in the direction of vorticity, $\boldsymbol{\Omega}\parallel \boldsymbol{k}$, the above scalar-vector modes become distinguishable from each other as the following:
	\begin{eqnarray}
	\omega^{\boldsymbol{\Omega}}_{3,4}&=&\pm \,c_s k \\
	\omega^{\boldsymbol{\Omega}}_{5,6}&=&\pm\, \Omega.\label{}
	\end{eqnarray}
	Clearly the modes $5,6$ are two standing vortex modes.
	In another limit when $\boldsymbol{\Omega}\perp \boldsymbol{k}$, we just obtain two sound waves gapped out by the background vorticity:
	\begin{eqnarray}
	\omega^{\boldsymbol{\Omega}}_{3,4}&=&\pm\,\sqrt{c_s^2 k^2 +\Omega_L^2}\,\,\,\,\rightarrow\,\,\,v^{\boldsymbol{\Omega}}_{3,4}=\pm \frac{c_s }{\sqrt{1+\frac{\Omega_L^2}{c_s^2 k^2}}}\\
	\omega^{\boldsymbol{\Omega}}_{5,6}&=&0.
	\end{eqnarray}
	The only modes propagating in transverse directions are $\omega_{3,4}$, the Coriolis-Sound waves, analogous to the magnetosonic waves \eqref{magnetosonic} in presence of transverse magnetic field.  
   Note that the anomaly effects can not be detected in directions transverse to the vorticity.

	In summary, when the wave vector is neither parallel nor transverse to the vorticity, the four scalar-vector modes become mixed Sound-Coriolis modes which also disperse when propagate.
	
		Before ending this section let us separate the new results of the paper in this part from their well-known counterpart in the literature. To our knowledge CVHWs \eqref{CVW-omega} and their corresponding eigenvectors \eqref{eigen_vec_omega_CVW} and \eqref{eigenvec3456Omega} were not found in previous studies.

	\section{Rotating QCD Fluid Coupled to Magnetic Field}
	\label{section_mixed}
	In this section we consider the general case in which the QCD fluid is either rotating and coupled to an external magnetic field. The associated results are lengthy and complicated, so we just limit ourself to write the hydrodynamic eigenmodes formally with a number of coefficients given in the related Appendix.

	\subsection{Equations of Motion Linearized}
	The thermodynamic equilibrium state of the fluid may be given by 
	\begin{equation}
	\begin{split}
	& u^{\mu}=\left(1,\frac{}{}\boldsymbol{\Omega} \times \boldsymbol{x}\right)\,\,\,\,\,\,\,\,\,\,\,\,\,\,\,\Omega\, r \ll 1,\\
	\,\,& T=Const.,\,\,\mu=Const.,\,\,\mu_5=Const.\\
	&\boldsymbol{B}=Const.
	\end{split}
	\end{equation}
	If we slightly perturb the above state as
	\begin{equation}
	\phi_{a}+\delta \phi_a=\left(T+\delta T, \,\frac{}{}\boldsymbol{0}+\boldsymbol{\pi},\,\mu+\delta \mu,\,\mu_5+\delta \mu_5\right),
	\end{equation}
	the linearized equations of motion take the following form
	\begin{equation}\label{eq:lin_mat}
	M^{B \Omega}_{ab}(\boldsymbol{k} , \omega)  \delta \phi_a (\boldsymbol{k} , \omega) = 0,
	\end{equation}
	with $M^{B \Omega}_{ab}$ given by  \eqref{matrix_BOmega} in Appendix \ref{MBOmega}.

	As in the previous two sections, two scalar modes together with four mixed scalar-vector modes constituted the full spectrum of the hydrodynamic excitations. As we will see, in the current subsection,  the scalar sector include the mixed CMWHWs, while in the scalar-vector sector there exist mixed Sound-Alfv\'en-Coriolis waves.
	\subsection{Hydro Modes}
	The dispersion relations of the two scalar modes, namely the CMVHWs, in this case are as the following
	\begin{equation}\label{CMWmixed}
	\begin{split}
	\omega^{\boldsymbol{B \Omega}}_{1,2}&=\,-\frac{1}{\mathcal{E}}\left(\mathcal{A}_1\frac{}{}\boldsymbol{B} \cdot  \boldsymbol{k}+\mathcal{A}_3\,\,\boldsymbol{\Omega} \cdot \boldsymbol{k}\right)\\
	&\pm\frac{1}{\mathcal{E}}\sqrt{\left(\mathcal{A}_1\frac{}{}\boldsymbol{B} \cdot  \boldsymbol{k}+\mathcal{A}_3\,\,\boldsymbol{\Omega} \cdot  \boldsymbol{k}\right)^2-\mathcal{E}\left(\mathcal{A}_1\big(\boldsymbol{B} \cdot \boldsymbol{k})^2 +\mathcal{A}_5(\boldsymbol{B} \cdot \boldsymbol{k})  \,\,(\boldsymbol{\Omega} \cdot \boldsymbol{k})+ \frac{}{}\mathcal{A}_4\,\,(\boldsymbol{\Omega} \cdot \boldsymbol{k})^2\right)}+O(k^3)
	\end{split}
	\end{equation}
	where the new anomaly expression is 
	\begin{equation}\label{z_1z_2}
	\mathcal{A}_5= \mathcal{C}^2 z_1+\, \mathcal{C D}\,T\,z_2
	\end{equation}
	with $z_1$ and $z_2$ given in Appendix
	\ref{z}.	
	These are in fact two waves with different velocities. In the non-chiral limit where $\mathcal{A}_1$ and $\mathcal{A}_3$ vanish, the velocities of two modes become the same.

	In the case of the scalar-vector modes, the dispersion relations are so complicated. We first give the dispersion relation of each mode at zero order of hydrodynamic constitutive currents:
	\begin{equation}
		\omega^{\boldsymbol{B \Omega}(1)}_{3,4,5,6}=\frac{\pm 1}{\sqrt{2}}\sqrt{\left(\boldsymbol{B}\frac{n}{w}+\boldsymbol{\Omega}\right)^2+c_s^2 k^2\pm\sqrt{\left(\left(\boldsymbol{B}\frac{n}{w}+\boldsymbol{\Omega}\right)^2+c_s^2 k^2\right)^2-4c_s^2\left(\boldsymbol{k} \cdot \boldsymbol{B}\frac{n}{w}+\boldsymbol{k} \cdot \boldsymbol{\Omega}\right)^2}}.
	\end{equation}
	By use of the above four zero order expressions, one may write the dispersion relations up to first order for $i=3,4,5,6$ as $\omega^{\boldsymbol{B \Omega}}_i=	\omega_i^{\boldsymbol{B \Omega}(1)}+		\omega_i^{\boldsymbol{B \Omega}(2)}$ with
	\begin{equation}\label{mixedmixed}
		\omega_{i}^{\boldsymbol{B \Omega}(2)}=-\frac{1}{\mathcal{E}}\,\,\frac{\left(\displaystyle\sum_{j} \frac{}{}a_j \mathbf{s}_j\right)\big(\omega_i^{\boldsymbol{B \Omega}(1)}\big)^4+\left(i \frac{}{}b \,\mathbf{s}_7\right)\big(\omega_i^{\boldsymbol{B \Omega}(1)}\big)^3+\left(\displaystyle\sum_{j,k}c_{j,k} \mathbf{s}_j \mathbf{s}_k\right)\big(\omega_i^{\boldsymbol{B \Omega}(1)}\big)^2+\,\displaystyle\sum_{j,k,l}\frac{}{}d_{j,k,l}  \mathbf{s}_j \mathbf{s}_k \mathbf{s}_l}{3\big(\omega_i^{\boldsymbol{B \Omega}(1)}\big)^4-\,2\left(k^2 c_s^2+\big(\boldsymbol{\Omega}+\frac{n}{w}\boldsymbol{B}\big)^2\right)\big(\omega_i^{\boldsymbol{B \Omega}(1)}\big)^2+\,c_s^2\left(\boldsymbol{\Omega}\cdot \boldsymbol{k}+\frac{n}{w}\boldsymbol{B}\cdot \boldsymbol{k}\right)^2}
	\end{equation}
	In the equation (\ref{mixedmixed}), $\{\mathbf{s}_i\}$ is the set of scalars made out of three independent vectors $\boldsymbol{k}$, $\boldsymbol{B}$ and $\boldsymbol{\Omega}$
	\begin{equation}
	\begin{split}
	\mathbf{s}_1&=\boldsymbol{k}^2,\,\,\,\,\,\,\,\,\,\,\,\,\,\,\,\\
	\mathbf{s}_2&=\boldsymbol{B}^2,\,\,\,\,\,\,\,\,\,\,\,\,\,\,\,\,\,\,\,\,\,\,\,\,	\mathbf{s}_3=\boldsymbol{B}\cdot\boldsymbol{k},\,\,\,\,\,\,\,\,\,\,\,\,\,\,\,\\
	\,\mathbf{s}_4&=\boldsymbol{\Omega}^2\,\,\,\,\,\,\,\,\,\,\,\,\,\,\,\,\,\,\,\,\,\,\,\,\,\,\,\,\mathbf{s}_5=\boldsymbol{\Omega}\cdot\boldsymbol{k},\\				\mathbf{s}_6&=\boldsymbol{B}\cdot\boldsymbol{\Omega},\,\,\,\,\,\,\,\,\,\,\,\,\,\,\,\,\,\,\,\,\,\,\mathbf{s}_7=\boldsymbol{k}\cdot\boldsymbol{B}\times\boldsymbol{\Omega}.
	\end{split}
	\end{equation}
	We have also defined a scalar $b$ (\ref{b}), a vector  $a_j$ (\ref{a_j}),  symmetric tensor $c_{j,k}$ (\ref{c_jk}) and tensor $d_{j,k,l}$ (\ref{d_jkl}) in the seven-dimensional space generated by the above scalars (see Appendix \ref{coef mix}). All these objects are in terms of the components of the susceptibility matrix and the anomaly coefficients.

\begin{figure}
			\centering
			\begin{tabular}{cccc}
\subfloat[][Mode with plus sign in \eqref{kpaBpaOmeg12}. \label{fig:sfig2}]{\includegraphics[width=7 cm]{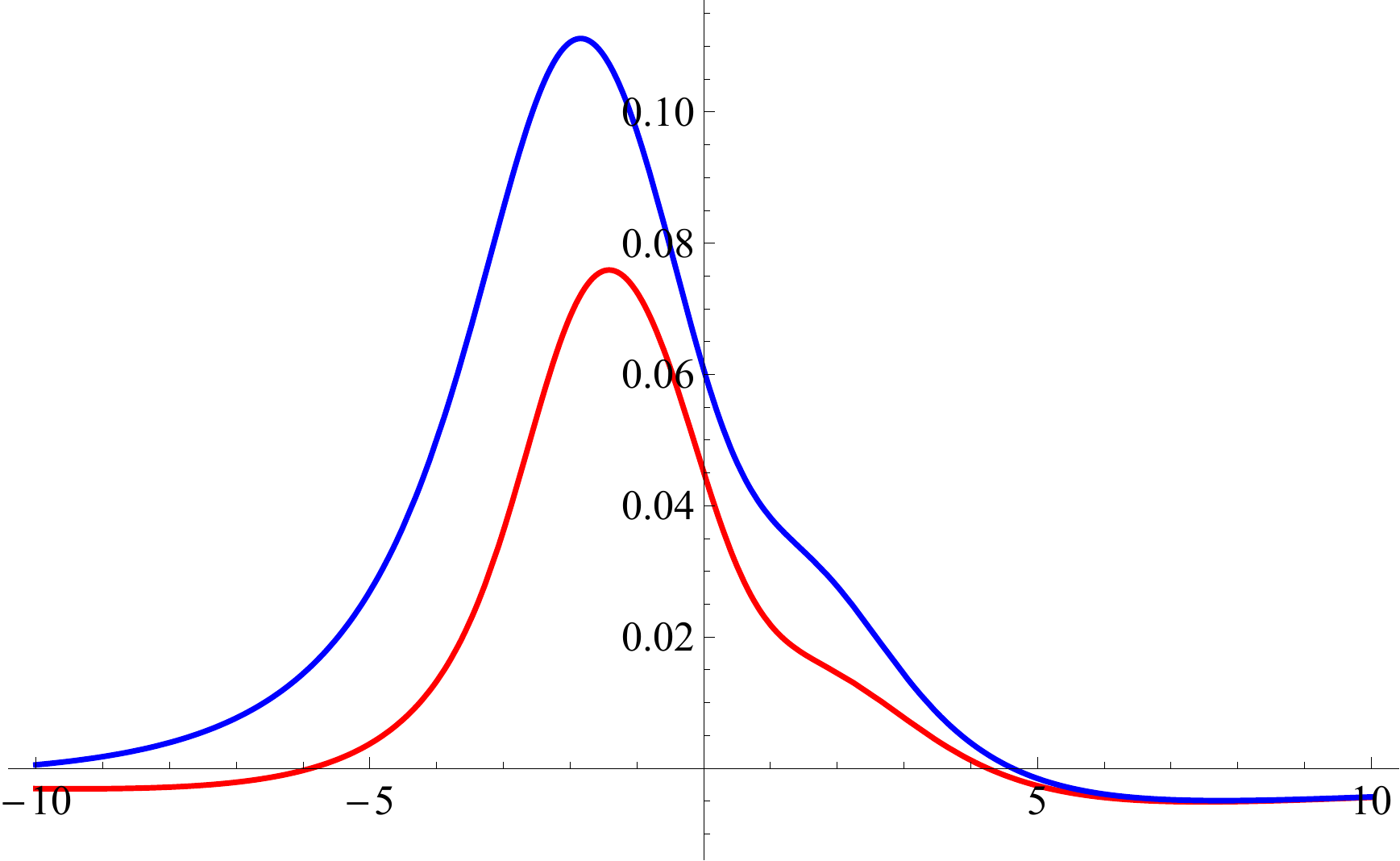}} &	
		\put(-30,0){$\mu/T$}
		\put(-110,130){$v_{CMW}$}
\subfloat[][Mode with minus sign in \eqref{kpaBpaOmeg12}. \label{fig:sfig1}]{%
  \includegraphics[width=7cm]{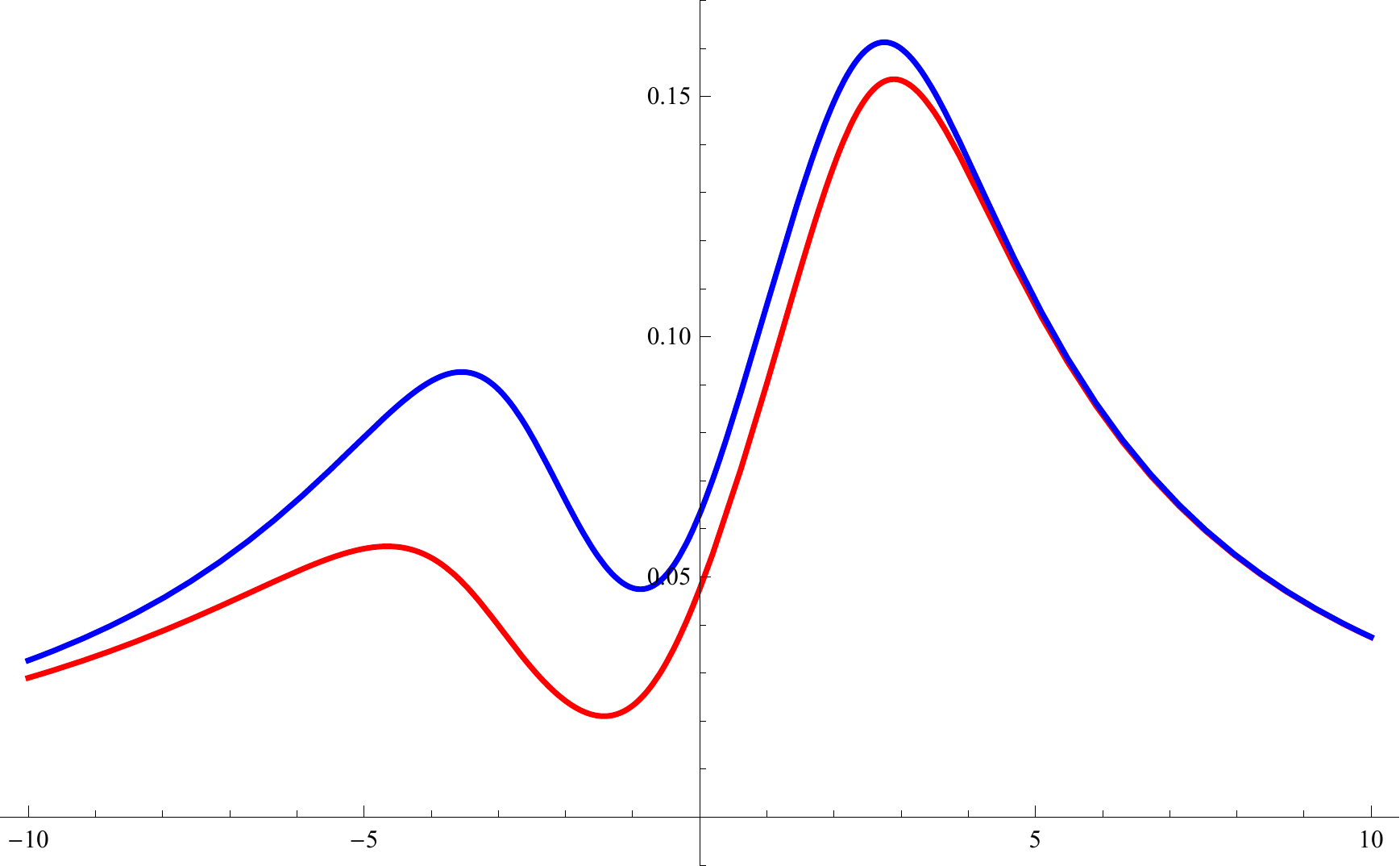}}&			
		\put(0,0){$\mu/T$}
		\put(-110,130){$v_{CMW}$}
		\end{tabular} 
		\caption{Velocity of mixed CMVHW, $v_{CMVHW}$ (red), compared to the sum of the velocities of the individual waves $v_{CMHW}+v_{CVHW}$ \,\,(blue). In both panels $\frac{\mu_5}{T}=2$ and $\frac{B}{T^2}=\frac{\Omega}{T}=0.5$.   We have used the equation of state \eqref{EoS}.}
		\label{fig3}
	\end{figure}

	Due to difficulties  in working with  the equation (\ref{mixedmixed}), from now on, we will focus on the special case wherein the magnetic field is parallel to the vorticity and study the propagation of waves along them. This case might be more relevant to the QCD fluid produced in heavy ion collisions.
	The dispersion relations of the modes in this case is as it follows
	\begin{eqnarray}\label{kpaBpaOmeg12}
	\omega^{\boldsymbol{B \Omega}}_{1,2}&=&\,-\frac{\mathcal{A}_1\,B + \mathcal{A}_3\,\Omega\pm\sqrt{\big( \mathcal{A}_1\;B + \mathcal{A}_3\,\Omega\big)^2-\mathcal{E}\left(\mathcal{A}_2\;B^2 +\mathcal{A}_5B\;\Omega+ \mathcal{A}_4\,\Omega^2\right)}}{\mathcal{E}}\;k\nonumber\\
	&&\\\label{kpaBpaOmeg34}
	\omega^{\boldsymbol{B \Omega}}_{3,4}&=&\pm \left(\frac{ n}{w}B+\Omega \right)-\left\{\mathcal{C}\, 
	\left(\mu \mu_5-\frac{n \mu_5}{3w}\left(3\mu^2+\mu_5^2\right)\right)-\mathcal{D}\,\frac{n \mu_5}{w}T^2\right\}\,\frac{B}{w}\,k\\
	\label{kpaBpaOmeg56}
	\omega^{\boldsymbol{B \Omega}}_{5,6}&=&\pm k c_s.
	\end{eqnarray}
	What we observe in the scalar sector is the existence of two CMVHWs. In figure \eqref{fig3}, we have plotted the velocities of these waves in two separate panels, the mode with plus sign in panel a and the mode with minus sign in panel b. 
	In each panel we have also plotted with a blue curve the following quantity:
	\begin{equation}\label{vsum}
	v_{sum}:=v_{CMHW}+v_{CVHW}=\,v_{CMVHW}|_{\boldsymbol{\Omega=0}}+v_{CMVHW}|_{\boldsymbol{B=0}}
	\end{equation}

	As the first point in the figure, the two CMVHWs neither always propagate  in the same direction nor have the same velocities.  More interestingly, one clearly sees that in general
	\begin{equation}
	v_{CMVHW}\, \ne \, v_{sum}.
	\end{equation}    
	It is simple to show that one of the situations in which $	v_{CMVHW}$ equates with $v_{sum}$ is within the quark gluon plasma produced in heavy ion collisions. In the latter case, $\mu_5=0$ and is equivalent to the limit $\frac{\mu}{\mu_5}\rightarrow\infty$. As it can be observed in Figure \eqref{fig3}, in this limit $\frac{v_{CMVHW}}{v_{sum}}\rightarrow1$.

In the case of modes \eqref{kpaBpaOmeg34} and \eqref{kpaBpaOmeg56}, one observes an interesting separation between the sound modes and the CAWs.  CAWs in this case are pure and  their propagation  is also accompanied with two oppositely polarizing vortices.

Before ending this subsection let us emphasize that to our knowledge	\textbf{all of the results} in this subsection are novel and have not been found in previous studies.
	
	\section{Phenomenology}
	\label{pheno}
	\subsection{ Application to Quark-Gluon-Plasma}
In this part, we want to apply the new results found in this paper to a real QCD-type fluid case, namely the QCD fluid produced in heavy ion collision experiments.
	It has been understood that the quark gluon plasma produced in a heavy ion collision is initially non-chiral, i.e. $\mu_5=0$.
	In this limit, we have $\beta_3=\alpha_3=0$ and the susceptibility matrix takes the following form:
	\begin{equation*}
	\tilde{\chi}=\,[\tilde{\chi}_{ij}]=\,	\begin{bmatrix}
	\,\,\,\frac{1}{c_s^2 T}(w- \mu n)&\,\,\,\,\,\,\,\frac{n}{c_s^2}\,\,\,\,\,\,\,\,0\,\,\,\\
	\frac{1}{T}\left(\frac{n}{c_s^2}-\chi \mu\right)&\,\,\,\,\,\,\chi\,\,\,\,\,\,\,\,\,\,0\,\,\,\\
	0&\,\,\,\,\,\,\,0\,\,\,\,\,\,\,\,\,\,\chi\,\,\,\\
	\end{bmatrix}.
	\end{equation*}

	As as result, the hydrodynamic modes given in (\ref{kpaBpaOmeg12}), (\ref{kpaBpaOmeg34}) and (\ref{kpaBpaOmeg56}) simplify to
	\begin{eqnarray}\label{most_important}
	\omega_{1,2}^{QGP}
&=&\pm\,\mathcal{C}\frac{Bk}{\chi }\sqrt{\frac{ \tilde{\chi}_{11}\tilde{\chi}_{22}}{\tilde{\chi}_{11}\tilde{\chi}_{22}-\tilde{\chi}_{12}\tilde{\chi}_{21}}}\pm\mathcal{C}\frac{\Omega \mu k}{\chi }\sqrt{\frac{\tilde{\chi}_{11}- \frac{\mathcal{D}T}{\mathcal{C}\mu}\tilde{\chi}_{12}}{\tilde{\chi}_{11}\tilde{\chi}_{22}-\tilde{\chi}_{12}\tilde{\chi}_{21}}}\\ \label{cawQGP}
	\omega_{3,4}^{QGP}&=&\pm \left(\Omega_{L}+\frac{}{}\Omega \right)\\
	\omega_{5,6}^{QGP}&=&\pm  c_s k.
	\end{eqnarray}
	Among the three equations given above, \eqref{most_important} is one of our main results regarding the QGP which deserves more explanations. To proceed, we first compute \eqref{vsum} for the current case, namely $v^{QGP}_{sum}$. A simple calculation shows that $v^{QGP}_{sum}$ is exactly the same as the velocity of mixed CMVHWs obtained from \eqref{most_important}:
	\begin{equation}
	v^{QGP}_{CMVHW}\, = \, v^{QGP}_{sum}.
	\end{equation}    
	This result means that $\mu_5=0$ case is an especial case in which CMHW and CVHW linearly mix to make CMVHWs; remember that  we showed in general, when $\mu\ne0$ and $\mu_5\ne0$, this equality does not continue to hold.
	 
	Let us recall that the expressions given in front of the square roots in \eqref{most_important} are nothing but the familiar chiral magnetic and chiral vortical waves found in \cite{Kharzeev:2010gd} and \cite{Jiang:2015cva} with assuming the energy and momentum perturbations being turned off.
	In \cite{Burnier:2011bf,Jiang:2015cva}, the induction of an electric quadrapole moment or equivalently the observation of difference between the elliptic flow of negative and  positive charged hadrons in QGP has been pointed out as the sign for the propagation of such waves, with the effect of CVW being weaker than that of CMW. Our computations show that
	even in presence of energy and momentum perturbations the quadrapole moment would be induced too,  while due to appearance of the square root expressions in \eqref{most_important}, the effect might be predicted slightly  different compared to \cite{Burnier:2011bf,Jiang:2015cva}.
	
	Another point with the hydro modes in QGP is that CAWs do not propagate in the plasma (see \eqref{cawQGP}) and the only propagating waves in the vector sector are two ordinary sound waves.

	\subsection{Comment on Dissipation in Quark-Gluon-Plasma}
	In the whole of this paper up to now, we focused on the propagation of waves in QCD-type chiral fluids  in the absence of dissipation. Considering the dissipative effects makes the computations extremely
	complicate and it would be so hard to extract interesting physics from that. In this subsection, we study such effects in the case of QGP fluid with this  simplifying point that there $\mu_5=0$. For more simplification, we neglect the effect of rotation in the plasma and just compute the hydrodynamic excitations in the magnetic field $B$. The set of six hydro modes are then given by:
	\begin{eqnarray}\label{CMHWdiss}
\omega_{1,2}^{QGP-diss}&=&	\frac{-i\,\sigma\,k^2}{2T}\,\left(\frac{ T \tilde{\chi}_{11}-\mu \tilde{\chi}_{12} }{\tilde{\chi}_{11}\tilde{\chi}_{22}-\tilde{\chi}_{12}\tilde{\chi}_{21}}\right)\\ \nonumber
&\pm &\sqrt{\left(\frac{\mathcal{C}B k}{\chi}\right)^2\,\frac{ \tilde{\chi}_{11}\tilde{\chi}_{22}}{\tilde{\chi}_{11}\tilde{\chi}_{22}-\tilde{\chi}_{12}\tilde{\chi}_{21}}-\left(\frac{\sigma k^2}{2T}\right)^2\left(
	\frac{ T \tilde{\chi}_{11}+\mu \tilde{\chi}_{12} }{\tilde{\chi}_{11}\tilde{\chi}_{22}-\tilde{\chi}_{12}\tilde{\chi}_{21}}\right)^2-2\frac{\mathcal{C}B k}{\chi}\frac{i\sigma_5 k^2 }{2T}\left(
	\frac{ T \tilde{\chi}_{11}+\mu \tilde{\chi}_{12} }{\tilde{\chi}_{11}\tilde{\chi}_{22}-\tilde{\chi}_{12}\tilde{\chi}_{21}}\right)}\\
\omega_{3,4}^{QGP-diss}&=&\pm \frac{n B}{w}	-i\,\frac{k^2 \eta + B^2 \sigma}{w}\,\\\label{sounddiss}
\omega_{5,6}^{QGP-diss}&=&\pm  c_s k- \frac{2 i k^2 \eta}{3 w}.
	\end{eqnarray}
	Obviously, the effect of the vector and chiral conductivities just appear in the first two modes. It is simple to see that when $\sigma=\sigma_5=0$, \eqref{CMHWdiss} gives the CMHWs in the first term of \eqref{most_important}. These two modes are the CMHWs which dissipate due to diffusion of vector and chiral charges. They are  in fact \textbf{dissipative} CMHWs. The next two modes, namely $\omega_{3,4}^{QGP-diss}$ are two oppositely circulating standing vortices which dissipate due to transverse shear effects as well as ohmic effects induced by the magnetic field. The same modes had been previously observed in 
	\cite{Abbasi:2015saa} in a chiral fluid with just one single chirality. Finally, the last two modes \eqref{sounddiss} are the  sound modes dissipating by the momentum diffusion in the transverse directions.

	\section{Conclusion and Outlook}
	\label{conclu}
	As a main part in this paper we have found the hydrodynamic excitations in a fluid carrying both vector and axial charges. Neglecting the dissipative effects, none of these excitations are entropy producing; they are  either adiabatic or anomalous waves in the fluid. In  the latter case, the chiral transport may be observed in the fluid when fluid is coupled to an external magnetic field or is rotating around an axis. 
	
	In this paper, we have considered a general case in which the fluid is in presence of a constant magnetic field $\boldsymbol{B}$ and simultaneously is rotating with a constant vorticity $\boldsymbol{\Omega}$.  It has been shown that the full spectrum of the collective excitations constitute six modes in general; two of them are the coherent perturbations of the scalar currents, namely  $J^{\mu}_{E}(=u_{\nu}T^{\nu\mu}), J^{\mu}, J^{\mu}_{5}$, while another four modes are made out of perturbations of all six scalar and vector currents.
	
	The scalar modes are the mixture of  CMHW and CVHW. 
	There is an interesting point about equation \eqref{kpaBpaOmeg12}. 
	We have found that $v_{CMVHW}$   is  actually a function of both $\mathcal{D}$ and $\mathcal{C}$. This suggests that by studying the effect of the chiral waves on  the final spectrum of the charged particle in QGP, it might be possible to investigate the presence of gravitational anomaly.   However, such observations require to do more precise experiments at higher energies compared to what currently is being done. 
	
	In the scalar-vector sector, we find four mixed Sound-Alfv\'en-Coriolis modes which are all dispersive in general. When $\boldsymbol{\Omega}=0$, these modes become the mixed Sound-Alfv\'en modes. While analogous \cite{Yamamoto:2015ria,Abbasi:2015saa} we have used the terminology of Alfv\'en here,  the Alfv\'en waves here are somewhat different from the standard Alfv\'en wave in magnetohydrodynamics. The main difference is that in the latter case the magnetic field has to be dynamical; in the former case however, we have shown chiral Alfv\'en waves propagate in presence of an external constant magnetic field.
	
	As mentioned above, the results in this paper have been found in presence of a non-dynamical magnetic field. It would be interesting to investigate how a dynamical magnetic field coupled to the flowing matter may affect on the nature of the excitations \footnote{See \cite{Giovannini:2013oga,Pandey:2016xpx,Sadooghi:2016ljd} for recent studies.}. To this end, one has to find the full spectrum of the chiral magnetohydrodynamics.
	We leave this issue for the future studies.

	It would be interesting to compute the full spectrum of the hydro modes in a QCD type fluid, microscopically. In the weak regime, using the recently developed chiral kinetic theory, one may extend the computations of \cite{Frenklakh:2016izv} to the case in which the axial and vector charge fluctuations are coupled to energy and momentum fluctuations. It should be noted that the chiral kinetic theory computations are  basically done in the Laboratory frame. It would be interesting to compare the results of the current paper in the Landau-Lifshitz frame with the results of Laboratory frame.     
	
	In another direction, it would be of more interest to find the spectrum of the  hydrodynamic excitations propagating on top of the expanding quark gluon plasma \footnote{The first attempt in this way, including CME however mostly numerically, was made in \cite{Taghavi:2013ena}.}. Recently, the authors of  \cite{Akamatsu:2016llw}  have studied the linear fluctuations around a Bjorken flow analytically although, neither they coupled the fluid to the magnetic field nor the chiral transport was considered in their work. It would be phenomenologically important to extend the subject of  \cite{Akamatsu:2016llw} to the chiral QCD case.  
	
	Apart from the quark gluon plasma, our results found in this paper may be applied to other phenomena in physics as well. A different place to explore is indeed the neutrino matter at the core of the supernova star wherein, a gas of noninteracting fermions is flowing \cite{Yamamoto:2015gzz}. It would be interesting to see how the velocities of the hydro waves change with the density there. We leave further study on the issue to our future work.

	\section*{Acknowledgements}
	We would like to thank Prof. Mohsen Alishahiha for encouragements and supporting the Larak-Particle-Pheno group. We would also like to thank M. Mohammadi Najafaabdi for reading the paper thoroughly and giving useful comments. We thank A. Akhavan. N.A. Would like to thank Prof. H. Arfaei for illuminating discussions on gravitational anomaly.  A.D would like to thank P.~V.~Buividovich and S.~N.~Valgushev for discussion. The work of A.D was supported by the S. Kowalevskaja award from the Alexander von Humboldt Foundation. We would like to thank Maxim Chernodub for discussion.

	\section{Appendix}
	\label{appendix}
	\subsection{Transforming from one thermo basis to another}
	\label{Appen thermo}
	Using the following thermodynamic relations, one can express the hydro modes in terms of the coherent excitatiions of a more physical set of varables, namely $\{\delta \epsilon, \boldsymbol{ }, \delta n,\delta n_5\}$:
	\begin{eqnarray}
	\delta T&=&\left(\frac{\partial T}{\partial \epsilon}\right) \delta \epsilon +\left(\frac{\partial T}{\partial n}\right) \delta n +\left(\frac{\partial T}{\partial n_5}\right) \delta n_5\\
	\delta \mu&=&\left(\frac{\partial \mu}{\partial \epsilon}\right) \delta \epsilon +\left(\frac{\partial \mu}{\partial n}\right) \delta n +\left(\frac{\partial \mu}{\partial n_5}\right) \delta n_5\\
	\delta \mu_5&=&\left(\frac{\partial \mu_5}{\partial \epsilon}\right) \delta \epsilon +\left(\frac{\partial \mu_5}{\partial n}\right) \delta n +\left(\frac{\partial \mu_5}{\partial n_5}\right) \delta n_5,
	\end{eqnarray}
	\subsection{Susceptibility Matrix and the Constraint Relations}
	\label{susceptibility matrix}
In order to express the dynamical fields $\epsilon$, $n$ and $n_5$ in terms of the variables (\ref{phia}) we consider the  Susceptibility Matrix as
\begin{equation}\label{susceptibility}
\tilde{\chi}=\begin{bmatrix}
\alpha_1&=\frac{\partial \epsilon}{\partial\small
	T},\,\,\,\,\,\,\,\,\alpha_2=\frac{\partial \epsilon}{\partial \mu},\,\,\,\,\,\,\,\,\alpha_3=\frac{\partial \epsilon}{\partial \mu_5}\\
\beta_1&=\frac{\partial n}{\partial T},\,\,\,\,\,\,\,\,\beta_2=\frac{\partial n}{\partial \mu},\,\,\,\,\,\,\,\,\beta_3=\frac{\partial n}{\partial \mu_5}\\
\gamma_1&=\frac{\partial n_5}{\partial T},\,\,\,\,\,\,\,\gamma_2=\frac{\partial n_5}{\partial \mu},\,\,\,\,\,\,\,\gamma_3=\frac{\partial n_5}{\partial \mu_5}\\
\end{bmatrix}.
\end{equation}
Let us recall that the elements of this matrix are not generally independent;  using the thermodynamics relations one simply show that:
\begin{eqnarray}
\beta_{1}&=&\frac{1}{c_s^2}\frac{n}{T}-\beta_2\frac{\mu}{T}-\gamma_2\frac{\mu_5}{T}\\
\beta_3&=&\gamma_2\\
\gamma_{1}&=&\frac{1}{c_s^2}\frac{n_5}{T}-\beta_3\frac{\mu}{T}-\gamma_3\frac{\mu_5}{T}.
\end{eqnarray}
\subsection{Matrix $M_{ab}^{B \Omega}$}
\label{MBOmega}
	The matrix $M_{ab}^{B \Omega}$ is given by:
\begin{equation}\label{matrix_BOmega}
\begin{bmatrix}
- i \alpha_1\omega & i k_j & -i \alpha_2 \omega & -i \alpha_3 \omega \\
& & & \\
i\alpha_1  v_s^2 k^i & -i\omega \delta^i_j -  \epsilon^i \,_{jl}\Omega^l  - \frac{\bar{n}}{\bar{w}} \epsilon^i \,_{jl}B^l&  i\alpha_2  v_s^2 k^i & i\alpha_3  v_s^2 k^i \\
& -i \frac{\xi}{2\bar{w}} \left(\boldsymbol{B} \cdot \boldsymbol{k} \delta^i_j - B_j k^i \right) & +\left(\frac{\partial \xi}{\partial \mu}\right)\underline{(\boldsymbol{B}\times \boldsymbol{\Omega})^i} & +\left(\frac{\partial \xi}{\partial \mu_5}\right)\underline{(\boldsymbol{B}\times \boldsymbol{\Omega})^i} \\
& & & \\
& & & \\
-i \beta_1 \omega+\left(\frac{\partial \xi}{\partial T}\right)  i \boldsymbol{\Omega} \cdot \boldsymbol{k} & \frac{\bar{n}}{\bar{w}} i k_j - \frac{2\xi}{\bar{w}} i\omega \Omega_j & -i\beta_2\omega +  \left(\frac{\partial \xi}{\partial \mu}\right) i \boldsymbol{\Omega} \cdot \boldsymbol{k} & -i \beta_3 \omega+ \left(\frac{\partial \xi}{\partial \mu_5}\right) i \boldsymbol{\Omega} \cdot \boldsymbol{k}\\
+\left(\frac{\partial \xi_B}{\partial T}\right)  i \boldsymbol{B} \cdot \boldsymbol{k}	& - \frac{\xi_{B}}{\bar{w}} i\omega B_j- \frac{\xi_{ B}}{\bar{w}} \underline{(\boldsymbol{B}\times \boldsymbol{\Omega})_j}& +\left(\frac{\partial \xi_B}{\partial \mu}\right) i \boldsymbol{B} \cdot \boldsymbol{k}& + \left(\frac{\partial \xi_B}{\partial \mu_5}\right) i \boldsymbol{B} \cdot \boldsymbol{k}  \\
& & & \\
& & & \\
-i \gamma_1 \omega+\left(\frac{\partial \xi_{5 }}{\partial T}\right)  i \boldsymbol{\Omega} \cdot \boldsymbol{k} & \frac{\bar{n_5}}{\bar{w}} i k_j - \frac{2\xi_{5}}{\bar{w}} i\omega \Omega_j &  -i \gamma_2 \omega+ \left(\frac{\partial \xi_{5}}{\partial \mu}\right) i \boldsymbol{\Omega} \cdot \boldsymbol{k} & - i\gamma_3  \omega+ \left(\frac{\partial \xi_{5 }}{\partial \mu_5}\right) i \boldsymbol{\Omega} \cdot \boldsymbol{k}
\\ +\left(\frac{\partial \xi_{5 B}}{\partial T}\right)  i \boldsymbol{B}\cdot \boldsymbol{k}	&- \frac{\xi_{5 B}}{\bar{w}} i\omega B_j- \frac{\xi_{5 B}}{\bar{w}} \underline{(\boldsymbol{B}\times \boldsymbol{\Omega})_j} &+\left(\frac{\partial \xi_{5B}}{\partial \mu}\right) i \boldsymbol{B} \cdot \boldsymbol{k} &+ \left(\frac{\partial \xi_{5B}}{\partial \mu_5}\right) i \boldsymbol{B} \cdot \boldsymbol{k} \\
& & & 
\end{bmatrix}.
\end{equation}
An interesting point with this matrix is the appearance of the terms including both vorticity and the magnetic field. Although, these terms disappear when the magnetic field is parallel to the vorticity. 
	\subsection{$x_1,x_2$ coefficients}
	\label{x}
	The anomaly coefficients in the structure of $\mathcal{A}_3$ and $\mathcal{A}_4$ in \ref{A3A4} are given by the following five expressions. The first two, namely $x_1$ and $x_2$ are in the structure of $\mathcal{A}_3$:
\newline
\begin{longtable}{|l|c|}
	\hline
Coefficient	& Structure    \\
	\hline
	\,\,\,\,\,\,\,\,\,	$x_1$	\,\,\,\,\,\,\,&$\,T\Biggl(\alpha_{[2} \beta_{3]}+\frac{
		T n_5}{w}\alpha_{[2} \beta_{1]}-\frac{
		T n}{w}\alpha_{[2} \gamma_{1]}+\frac{
		n_5 \mu_5}{w}\alpha_{[3} \beta_{2]}-\frac{
		n \mu_5}{w}\alpha_{[3} \gamma_{2]}+\frac{2 \mu_5 T}{w}\Biggr)$\\
	\hline
	&\\
	\,\,\,\,\,\,\,\,\,$x_2$	\,\,\,\,\,\,\,&$\frac{1}{ w} \left(n_5\,\alpha_{[2} \beta_{1]}-n\,\alpha_{[2} \gamma_{1]}\right)(\mu^2+\mu_5^2)+\frac{2\mu\mu_5}{w}(n_5\,\alpha_{[1} \beta_{3]}-n\,\alpha_{[1} \gamma_{3]})$\\
	&\,\,\,\,\,\,\,\,\,$- \mu\,\left(\alpha_{[1} \beta_{3]}+\frac{}{}\alpha_{[1} \gamma_{2]}\right)+ \mu_5\left(\alpha_{[1} \beta_{2]}+\frac{}{}\alpha_{[1} \gamma_{3]}\right)+\frac{2\mu_5}{w}\left(\mu^2+\frac{\mu_5^2}{3}\right)\mathcal{E} $\\
	\hline
\end{longtable}
	\subsection{$ y_1,y_2,y_3$ coefficients}
	\label{y}
		The anomaly coefficients  in the structure of $\mathcal{A}_4$  in \ref{A3A4}  are $y_1$, $y_2$ and $y_3$:
 	\newline
	\begin{longtable}{|l|c|}
		\hline
	Coefficient	&  Structure    \\
		\hline
		&\\
		\,\,\,\,\,\,\,\,\,			$	y_1$	\,\,\,\,\,\,\,\,\,	 &$4\,\alpha _1 \mu ^2\left(1-\frac{\mu  n}{w}\right)-4\,\alpha _1 \mu_5 ^2\left(1-\frac{\mu_5  n_5}{w}\right)+\frac{4\,\alpha _1 \mu  \mu _5}{w}(\mu_5 n- \mu n_5)$\\
		&$-\frac{4\,(\alpha _{[1} \beta _{3]}+\alpha _{[1} \gamma _{2]}) \mu  \mu _5}{w}\left(\mu^2+\frac{\mu_5^2}{3}\right)+\frac{4\,(\alpha _{[1} \beta _{2]}+\alpha _{[1} \gamma _{3]})  \mu _5^2}{w}\left(\mu^2+\frac{\mu_5^2}{3}\right)$\\
		&$+(\alpha_{[2}\beta_{1]}n_5-\alpha_{[2}\gamma_{1]}n)\frac{16 \mu ^2 \mu _5^3 }{3 w^2}+2(\alpha_{[1}\beta_{3]}n_5-\alpha_{[1}\gamma_{3]}n)\frac{4 \mu \mu_5^2 }{w^2}\left(\mu^2+\frac{\mu_5^2}{3}\right)$\\
		&$+(\alpha_{[2}\beta_{1]}n_5-\alpha_{[2}\gamma_{1]}n)\frac{4\,\mu _5 }{w^2}\left(\mu^4+\frac{\mu_5^4}{3}\right)+\frac{4\,\mu_5^2}{w^2}\left(\mu^2+\frac{\mu_5^2}{3}\right)^2\mathcal{E}$\\
		&\\
		\hline
		\,\,\,\,\,\,\,\,\,		$	y_2$	\,\,\,\,\,\,\,\,\,	&$ 4\,T^3 \Biggl(\frac{\alpha _2 n}{w}+(\alpha_{[1} \gamma_{2]}n-\alpha_{[1} \beta_{2]}n_5)\frac{\mu_5 T}{3 w^2}+2(\alpha_{[2} \gamma_{3]}n-\alpha_{[2} \beta_{3]}n_5)\frac{\mu_5^2}{ w^2}
		+\alpha_{[2} \beta_{3]}\frac{\mu _5 }{w}+\frac{\mu _5^2 T }{w^2}\mathcal{E}\Biggr) $\\
		\hline
		\,\,\,\,\,\,\,\,\,		$	y_3$	\,\,\,\,\,\,\,\,\,	&$4\,T\Biggl((\alpha _3 \mu _5-\alpha_2 \mu)\left(1-\frac{2 \mu _5 n_5}{w}\right)+\alpha _1 (\mu _5 n_5-\mu n)\frac{ T}{w}+\alpha _2 (\mu ^2 -\mu_5^2)\frac{n}{w}$\\
		&$+\left(\alpha_{[2} \beta_{3]}\frac{\mu_5}{w}+\frac{2\mu_5^2 n}{w^2}\alpha_{[2} \gamma_{3]}-\frac{2\mu_5^2 n_5}{w^2}\alpha_{[2} \beta_{3]}\right)\left(\mu^2+\frac{\mu_5^2}{3}\right)+ (\alpha_{[1} \gamma_{2]}n-\alpha_{[1} \beta_{2]}n_5)\frac{4\mu^3 T}{3 w^2}$\\
		&$+\left((\alpha_{[1} \beta_{3]}\mu_5-\alpha_{[1} \beta_{2]}\mu)n_5+\frac{}{}(\alpha_{[1} \gamma_{2]}\mu-\alpha_{[1} \gamma_{3]}\mu_5)n\right)\frac{2 \mu  \mu _5  T}{w^2}$\\
		&$+\left((\alpha_{[1} \beta_{2]}\mu_5-\alpha_{[1} \gamma_{2]}\mu)+\frac{}{}(\alpha_{[1} \gamma_{3]}\mu_5-\alpha_{[1} \beta_{3]}\mu)\right)\frac{\mu _5 T}{w}
		+\frac{2 \mu _5^2 T}{w^2}\left(\mu^2+\frac{\mu_5^2}{3}\right)\mathcal{E}\Biggr)$\\
		\hline
	\end{longtable}
	\subsection{ $z_1,z_2$  coefficients}
	\label{z}
		The anomaly coefficients in the structure of $\mathcal{A}_5$ in \eqref{z_1z_2}, namely $z_1$ and $z_2$ are given by:
		\begin{longtable}{|l|c|}
			\hline
		Coefficient	&  Structure    \\
			\hline
			&\\
			\,\,\,\,\,\,\,\,\,		$	z_1$	\,\,\,\,\,\,\,\,\,& $2(3 \mu ^2 \mu _5^2+\frac{1}{3} \mu _5^4)(n_5\alpha_{[1} \beta_{3]}-n\frac{}{}\alpha_{[1} \gamma_{3]})-2(\frac{4}{3} \mu  \mu _5^3+2 \mu ^3 \mu _5)(n_5\alpha_{[1} \beta_{2]}-n\frac{}{}\alpha_{[1} \gamma_{2]})$\\
			& $-2w( \mu ^2 \mu _5 +\frac{1}{3} \mu _5^3 )(\alpha_{[1} \beta_{3]}+\frac{}{}\alpha_{[1} \gamma_{2]})+2\mu  \mu _5^2 w(\alpha_{[1} \beta_{2]}+\frac{}{}\alpha_{[1} \gamma_{3]})+4 \mu  \mu _5^2\left(\mu^2+\frac{\mu_5^2}{3}\right)\mathcal{E}$\\
			& $-4\alpha_1 \mu w(\mu n+\mu _5 n_5-w)$\\
			\hline
			& \\
			\,\,\,\,\,\,\,\,\,		$	z_2$	\,\,\,\,\,\,\,\,\,& $-4 \mu  \mu _5^2(n_5\alpha_{[2} \beta_{3]}-n\frac{}{}\alpha_{[2} \gamma_{3]})+2\mu  \mu _5 w\alpha_{[2} \gamma_{3]}$\\
			& $+2\mu_5^2(n_5\alpha_{[1} \beta_{3]}-n\frac{}{}\alpha_{[1} \gamma_{3]})-2\mu _5 T w(\alpha_{[1} \beta_{3]}+\frac{}{}\alpha_{[1} \gamma_{2]})
			-4 \mu  \mu _5 n_5 T(n_5\alpha_{[1} \beta_{2]}-n\frac{}{}\alpha_{[1} \gamma_{2]})$\\
			& $-2\left(\alpha _1 n T -\alpha_2(\mu  n+2 \mu _5 n_5-w)-\alpha _3 \mu _5 n\right)w +4 \mu  \mu _5^2 T \mathcal{E}$\\
			& \\
			\hline
		\end{longtable}
\subsection{$ b$ coefficient}
\label{b}
	\label{coef mix}
	The only scalar coefficient in \eqref{mixedmixed} is $b$ which given by
	\begin{longtable}{|l|c|}
		\hline
	\,\,\,\,	$b$\,\,\,\,& $2 \,\mathcal{C}\Biggl\{\frac{1}{2}(\mu_5\beta_{[1} \gamma_{3]}-\mu\beta_{[1} \gamma_{2]})	+\frac{n }{2w}\left(\beta_{[1} \gamma_{2]}-\frac{n}{ w}\alpha_{[1} \gamma_{2]}+\frac{n_5}{ w}\alpha_{[1} \beta_{2]}\right)(\mu^2+\mu_5^2)+\frac{\mu }{2w}(n	\alpha_{[1} \gamma_{2]}-n_5\alpha_{[1} \beta_{2]})$\\

	& $+\frac{\mu_5 }{w}\left(\frac{\mu n}{w}-\frac{1}{2}	\right)(n	\alpha_{[1} \gamma_{3]}-n_5\alpha_{[1} \beta_{3]})-\frac{\mu \mu_5 n}{w}\beta_{[1} \gamma_{3]}+\frac{\mu_5 n}{2w^2}\left(c_s^2-\frac{1}{2}\right)\left(\mu ^2 +\frac{\mu_5^2}{3}\right)\mathcal{E}+\frac{\mu \mu_5}{4 w}\mathcal{E}\Biggr\}$\\
		& $2\,\mathcal{D}T\Biggl\{\frac{n T }{2w}\left(\beta_{[1} \gamma_{2]}-\frac{n}{ w}\alpha_{[1} \gamma_{2]}+\frac{n_5}{ w}\alpha_{[1} \beta_{2]}\right)\,\,\,\,+\frac{n \mu_5}{w}\left(\beta_{[2} \gamma_{3]}-\frac{n}{ w}\alpha_{[2} \gamma_{3]}+\frac{n_5}{ w}\alpha_{[2} \beta_{3]}\right)+\frac{\mu _5 n T}{2 w^2}\left(c_s^2-\frac{1}{2}\right)\mathcal{E}\Biggr\}$
		\\
		\hline
	\end{longtable}
\subsection{$ a_j$ coefficients}
\label{a_j}
	The non-vinishing components of the vector $a_j$ in \eqref{mixedmixed} are $a_3$ and $a_5$ as the following:
		\begin{longtable}{|l|c|}
    	\hline
       	Coefficient	&  Structure    \\
		\hline
		\,\,\,\,\,\,\,\,\,		$	a_3$	\,\,\,\,\,\,\,\,\,& $\mathcal{C}\Biggl\{\frac{1}{2}(\alpha_{[3} \beta_{1]}+\alpha_{[2} \gamma_{1]})+\left(\frac{\mu _5 n_5}{2w} \alpha _{[1} \beta _{3]}-\frac{\mu  n_5}{2w} \alpha _{[1} \beta _{2]}+\frac{\mu  n}{2w} \alpha _{[1} \gamma _{2]}-\frac{\mu _5 n}{2w} \alpha _{[1} \gamma _{3]}\right)$\\
		& \,\,\,\,$-\frac{\mu _5 n}{ w^2}\left(\mu ^2 +\frac{\mu_5^2}{3}\right)\mathcal{E}+\frac{6 \mu \mu_5}{ w}\mathcal{E}\Bigg\}-\mathcal{D}\Bigg\{\frac{\mu _5 n T^2 }{ w^2}\Biggr\}\mathcal{E}$ \\
		& 	\\
		\hline
		& \\
		\,\,\,\,\,\,\,\,\,		$a_5$	\,\,\,\,\,\,\,\,\,& $2\,\mathcal{C}\Bigg\{\frac{1}{2}(-\mu  \alpha _{[1} \beta _{3]}+\mu _5 \alpha _{[1} \beta _{2]}-\mu  \alpha _{[1} \gamma _{2]}+\mu _5\alpha _{[1}  \gamma _{3]})+\frac{\mu_5}{w}\left(\mu ^2 +\frac{\mu_5^2}{3}\right)\mathcal{E}$\\
		& \,\,\,\,$+\left(\alpha _{[1} \gamma _{2]}\frac{n }{2 w}-\alpha _{[1} \beta _{2]}\frac{n_5 }{2 w}\right)(\mu ^2 +\mu_5 ^2 )+(\alpha _{[1} \beta _{3]}-\alpha _{[1} \gamma _{3]})\frac{\mu \mu_5 n_5 }{ w}\Biggr\}$\\
		& \,$+2\mathcal{D}T\Biggl\{\frac{1}{2} \alpha _{[2} \beta _{3]}+\frac{ \mu_5}{w}(\alpha _{[2} \gamma _{3]} n-\alpha _{[2} \beta _{3]} n_5)+\frac{  T}{2w}(n\alpha _{[1} \gamma _{2]} -n_5\alpha _{[1} \beta _{2]} )+\frac{\mu_5 T}{w}\mathcal{E}\Biggr\}$\\
		\hline
	\end{longtable}
	%
	%
	\subsection{$d_{j,k,l}$ coefficients}
	\label{d_jkl}
	The tensor $d_{j,k,l}$ in \eqref{mixedmixed} is a fully symmetric rank-$3$ tensor with the following non-vanishing components:
	\begin{longtable}{|l|c|}
	\hline
	Coefficient	& Structure   \\
	\hline
	\,\,\,\,\,\,\,\,\,	$
	d_{3,3,3}$	\,\,\,\,\,\,\,\,\,	& $\mathcal{C}\Biggl\{\frac{c_s^2 n^2}{2 w^3}\left(\mu n\alpha_{[1}\gamma_{2]}-\mu_5 n\alpha_{[1}\gamma_{3]}-\mu n_5\alpha_{[1}\beta_{2]}+\mu_5 n_5\alpha_{[1}\beta_{3]}\right)-\frac{c_s^2 n^2}{2 w^2}(\alpha_{[1} \beta_{3]}+\alpha_{[1} \gamma_{2]})+\frac{c_s^2\mu  \mu _5 n^2}{w^3}\mathcal{E}\Biggr\}$\\
	\hline
	\,\,\,\,\,\,\,\,\,		$d_{5,5,5}$ 	\,\,\,\,\,\,\,\,\,& $\mathcal{C}\Biggl\{c_s^2\mu_5 (\alpha_{[1} \beta_{2]}+\frac{}{}\alpha_{[1} \gamma_{3]})-c_s^2\mu(\alpha_{[1} \beta_{3]}+\frac{}{}\alpha_{[1} \gamma_{2]})+\frac{2\,c_s^2 \mu _5}{w}\left(\mu^2+\frac{\mu_5^2}{3}\right)$\\
	& \,\,\,\,\,\,\,\,\,\,\,\,\,\,\,\,\,$-\frac{c_s^2}{ w}(n_5\alpha_{[1} \beta_{2]}-n\frac{}{}\alpha_{[1} \gamma_{2]})(\mu^2+\mu_5^2)+\frac{2\,c_s^2 \mu  \mu _5}{w}(n_5\alpha_{[1} \beta_{3]}-n\frac{}{}\alpha_{[1} \gamma_{3]})\Biggr\}$\\
	& $+\mathcal{D}  T\Biggl\{-\frac{c_s^2 T}{ w}(n_5\alpha_{[1} \beta_{2]}-n\frac{}{}\alpha_{[1} \gamma_{2]})-\frac{2\,c_s^2 \mu _5}{w}(n_5\alpha_{[2} \beta_{3]}-n\frac{}{}\alpha_{[2} \gamma_{3]})+c_s^2\alpha_{[2} \beta_{3]}+\frac{2\,c_s^2 \mu _5 T}{w}\mathcal{E}\Biggr\}$\\
	\hline
		\,\,\,\,\,\,\,\,\,	$d_{3,3,5}$	\,\,\,\,\,\,\,\,\,& $\mathcal{C}\Biggl\{-\frac{c_s^2 n^2}{ w^3}(n_5\alpha_{[1} \beta_{2]}-n\frac{}{}\alpha_{[1} \gamma_{2]})( \mu ^2+ \mu _5^2)
	+\frac{2\,c_s^2 \mu  \mu _5 n^2}{w^3}(n_5\alpha_{[1} \beta_{3]}-n\frac{}{}\alpha_{[1} \gamma_{3]})$\\
		\,\,\,\,\,\,\,\,\,	$=d_{5,3,3}$	\,\,\,\,\,\,\,\,\,& $+\,\frac{c_s^2 n^2}{2 w^2}\left(\mu_5(\alpha_{[1} \beta_{2]}-\frac{}{}\alpha_{[1} \gamma_{3]})-\mu(\alpha_{[1} \beta_{3]}-\frac{}{}\alpha_{[1} \gamma_{2]})\right)+\frac{2\,c_s^2 \mu _5 n^2}{w^3}\left(\mu^2+\frac{\mu_5^2}{3}\right)\mathcal{E}$ \\
	\,\,\,\,\,\,\,\,\,	$=d_{3,5,3}$	\,\,\,\,\,\,\,\,\,	& $+\,\frac{c_s^2 n_5 n}{w^2}(\mu_5\alpha_{[1} \beta_{3]}-\mu\alpha_{[1} \beta_{2]})-\frac{c_s^2 n}{w}(\alpha_{[1} \beta_{3]}+\frac{}{}\alpha_{[1} \gamma_{2]})
	+\frac{ 2\,c_s^2 \mu  \mu _5 n}{w^2}\mathcal{E}\Biggr\}$\\
	& $-\mathcal{D}T\Biggl\{\frac{c_s^2 n^2 T}{ w^3}\left((n_5\alpha_{[1} \beta_{2]}-n\frac{}{}\alpha_{[1} \gamma_{2]})+4 \mu_5(n_5\alpha_{[2} \beta_{3]}-n\frac{}{}\alpha_{[2} \gamma_{3]})\right)
	-\frac{2\,c_s^2 n^2}{w^2}\alpha_{[2} \beta_{3]}-\frac{c_s^2\mu _5 n^2 T^2}{w^3}\mathcal{E}\Biggr\}.$\\
	\hline
	\end{longtable}
	%
\subsection{$ c_{j,k}$ coefficients}
\label{c_jk}
	The tensor $c_{j,k}$ in \eqref{mixedmixed} is a symmetric tensor with the following non-vanishing components:
\newline
\begin{longtable}{|l|c|}
	\hline
	Coefficient &  Structure    \\
	\hline
\,\,\,\,\,\, $
	c_{3,1}=	c_{1,3}$	\,\,\,\,\,\, & $\mathcal{C}\Biggl\{\frac{c_s^2}{2}(\alpha_{[1} \beta_{3]}+\alpha_{[1} \gamma_{2]})-\frac{c_s^2}{2 w}\left(\mu n\alpha_{[1}\gamma_{2]}-\mu_5 n\alpha_{[1}\gamma_{3]}-\mu n_5\alpha_{[1}\beta_{2]}+\mu_5 n_5\alpha_{[1}\beta_{3]}\right)$\\
	& \,\,\,\,\,$+\frac{c_s^2 \mu _5 n}{ w^2}\left(\mu^2+\frac{\mu_5^2}{3}\right)\mathcal{E}-\frac{3 c_s^2 \mu  \mu _5}{ w}\mathcal{E}\Biggr\}
	+\mathcal{D}  T\Biggl\{\frac{c_s^2 \mu _5 n T^2 }{ w^2}\mathcal{E}\Biggr\}$\\
	\hline
\,\,\,\,\,\,	$c_{5,2}=c_{2,5}$	\,\,\,\,\,\,& $2\,\mathcal{C}\Biggl\{\frac{ n}{2 w^2}\left((\mu n-\mu_5 n_5)\alpha_{[1}\beta_{3]}+(\mu n_5-\mu_5 n)\alpha_{[1}\beta_{2]}\right)-\frac{n^2}{2 w^2}\beta_{[1}\gamma_{2]}(\mu^2+\mu_5^2)+\frac{\mu  n}{2 w}\beta_{[1}\gamma_{2]}$\\
		& $-\frac{\mu _5 n}{2 w}\left(1-\frac{2\mu  n}{w}\right)\beta_{[1}\gamma_{3]}-\frac{\mu _5 n^2}{w^3}\left(\frac{c_s^2}{2}+\frac{5}{4}\right)\left(\mu^2+\frac{\mu_5^2}{3}\right)\mathcal{E}+\frac{\mu  \mu _5 n}{4 w^2}\mathcal{E}\Biggr\}$\\
	& \,$+2\mathcal{D} T\Biggl\{-\frac{n^2}{2w^2}\left(2\mu_5\beta_{[2}\gamma_{3]}+\alpha_{[2}\beta_{3]}+T\beta_{[1}\gamma_{2]}\right)-\frac{\mu _5 n^2 T}{w^3}\left(\frac{c_s^2}{2}+\frac{5}{2}\right)\mathcal{E}\Biggr\}$
	\\
	\hline
\,\,\,\,\,\,$c_{5,6}=	c_{6,5}$\,\,\,\,\,\,	& $\mathcal{C}\Biggl\{(\mu\beta_{[1}\gamma_{2]}-\mu_5\beta_{[1}\gamma_{3]})-\frac{ n}{ w}\left(\beta_{[1}\gamma_{2]}+\frac{n}{w}\alpha _{[1} \gamma _{2]}-\frac{n_5}{w}\alpha _{[1} \beta _{2]}\right)(\mu^2+\mu_5^2)$\\
	& $+\frac{2\,\mu  \mu _5 n}{w}\left(\beta_{[1}\gamma_{3]}+\frac{n}{w}\alpha _{[1} \gamma _{3]}-\frac{n_5}{w}\alpha _{[1} \beta _{3]}\right)+\frac{2\,n}{w}(\mu\alpha_{[1}\beta_{3]}-\mu_5 \alpha_{[1}\beta_{2]})+\frac{\mu  \mu _5 }{2 w}\mathcal{E}$\\
	& $+\frac{ 1}{ w}\left(\mu n\alpha_{[1}\gamma_{2]}-\mu_5 n\alpha_{[1}\gamma_{3]}+\mu n_5\alpha_{[1}\beta_{2]}-\mu_5 n_5\alpha_{[1}\beta_{3]}\right)
	-\frac{ \mu _5 n}{ w^2}\left(c_s^2+\frac{9}{2}\right)\left(\mu^2+\frac{\mu_5^2}{3}\right)\mathcal{E}\Biggr\}$\\
	& $+2\,\mathcal{D}T\Biggl\{-\frac{n}{w}\alpha _{[2} \beta _{3]}-\frac{n T}{2 w}\left(\beta_{[1}\gamma_{2]}+\frac{n}{w}\alpha _{[1} \gamma _{2]}-\frac{n_5}{w}\alpha _{[1} \beta _{2]}\right)$\\
	& $-\frac{\mu _5 n}{w}\left(\beta _{[2} \gamma _{3]}+\frac{n}{w}\alpha _{[2} \gamma _{3]} -\frac{n_5}{w}\alpha _{[2} \beta _{3]} \right)-\frac{\mu _5 n T}{ w^2}\left(\frac{c_s^2}{2}+\frac{9}{4}\right)\mathcal{E}\Biggr\}
	$\\
	\hline
 \,\,\,\,\,\,$ c_{5,4}=c_{4,5}$ \,\,\,\,\,\,	& $\mathcal{C}\Biggl\{\mu(\alpha_{[1} \beta_{3]}+\frac{}{}\alpha_{[1} \gamma_{2]})-\mu_5(\alpha_{[1} \beta_{2]}+\frac{}{}\alpha_{[1} \gamma_{3]})\Biggr\}$\\
		& $+\mathcal{D} T\Biggl\{\frac{1}{w}(n_5\alpha_{[1} \beta_{2]}-n\frac{}{}\alpha_{[1} \gamma_{2]})(\mu^2+\mu_5^2)-\frac{2\,\mu  \mu _5 }{w}(n_5\alpha_{[1} \beta_{3]}-n\frac{}{}\alpha_{[1} \gamma_{3]})
		-\frac{2\,\mu_5 }{w}\left(\mu^2+\frac{\mu_5^2}{3}\right)\mathcal{E}\Biggr\}$\\
		\hline
	\,\,\,\,\,\,	$c_{5,1}= 	c_{1,5}$ \,\,\,\,\,\,& $\mathcal{C}\Biggl\{\mu c_s^2(\alpha_{[1} \beta_{3]}+\frac{}{}\alpha_{[1} \gamma_{2]})- \mu_5 c_s^2(\alpha_{[1} \beta_{2]}+\frac{}{}\alpha_{[1} \gamma_{3]})-\frac{2\,c_s^2\mu_5 }{w}\left(\mu^2+\frac{\mu_5^2}{3}\right)\mathcal{E}$\\
		& $+\frac{c_s^2}{ w}(n_5\alpha_{[1} \beta_{2]}-n\frac{}{}\alpha_{[1} \gamma_{2]})(\mu^2+\mu_5^2)
		-\frac{2\,c_s^2\mu  \mu _5 }{w}(n_5\alpha_{[1} \beta_{3]}-n\frac{}{}\alpha_{[1} \gamma_{3]})\Biggr\}$\\
		&$+\mathcal{D} T\Biggl\{\frac{c_s^2 T}{ w}(n_5\alpha_{[1} \beta_{2]}-n\frac{}{}\alpha_{[1} \gamma_{2]})+\frac{2\,c_s^2\mu _5}{w}(n_5\alpha_{[2} \beta_{3]}-n\frac{}{}\alpha_{[2} \gamma_{3]})
		-c_s^2 \alpha_{[2} \beta_{3]}-\frac{2\,c_s^2 \mu _5 T}{w}\mathcal{E}\Biggr\}$\\
		\hline
	 \,\,\,\,\,\,	$c_{3,6}=c_{6,3}$ \,\,\,\,\,\, & $\mathcal{C}\Biggl\{\frac{ n^2} { w^2}\left(\beta_{[1}\gamma_{2]}-\frac{n}{w}\alpha _{[1} \gamma _{2]}+\frac{n_5}{w}\alpha _{[1} \beta _{2]}\right)(\mu ^2+\mu ^2_5)+\frac{ n}{2 w^2}\left(\mu n\alpha_{[1}\gamma_{2]}-\mu_5 n\alpha_{[1}\gamma_{3]}+\mu n_5\alpha_{[1}\beta_{2]}-\mu_5 n_5\alpha_{[1}\beta_{3]}\right)$ \\
		& $-\frac{2\,\mu  \mu _5 n^2}{w^2}\left(\beta _{[1} \gamma _{3]}-\frac{n}{w}\alpha _{[1} \gamma _{3]} +\frac{n_5}{w}\alpha _{[1} \beta _{3]} \right)+\frac{n }{2w}(\alpha _{[1} \beta _{3]}+\alpha _{[1} \gamma _{2]})-\frac{9 \mu  \mu _5 n}{4 w^2}\mathcal{E}+\frac{ n }{ w}(\mu_5\beta _{[1} \gamma _{3]}-\mu\beta _{[1} \gamma _{2]})$ \\
		& $+\frac{ \mu _5 n^2}{ w^3}\left(c_s^2+\frac{1}{2}\right)\left(\mu^2+\frac{\mu_5^2}{3}\right)\mathcal{E}\Biggr\}
		+\mathcal{D} T\Biggl\{\frac{2\,\mu _5 n^2}{w^2}\left(\beta _{[2} \gamma _{3]}-\frac{n}{w}\alpha _{[2} \gamma _{3]} +\frac{n_5}{w}\alpha _{[2} \beta _{3]} \right)+\frac{\mu _5 n^2 T^2}{ w^3}\left(c_s^2+\frac{1}{2}\right)\mathcal{E}$\\
		& $+\frac{n^2 T}{ w^2}\left(\beta _{[1} \gamma _{2]}-\frac{n}{w}\alpha _{[1} \gamma _{2]} +\frac{n_5}{w}\alpha _{[1} \beta _{2]} \right)\Biggr\}$
\\		\hline
		 \,\,\,\,\,\,$c_{2,3}=c_{3,2}$ \,\,\,\,\,\, & $\mathcal{C}\Biggl\{\frac{n^2}{2 w^2}(\alpha_{[1} \beta_{3]}+\alpha_{[1} \gamma_{2]})-\frac{\mu  \mu _5 n^2}{w^3}\mathcal{E}$\\
		& $-\frac{ n^2}{2 w^3}\left(\mu n\alpha_{[1}\gamma_{2]}-\mu_5 n\alpha_{[1}\gamma_{3]}-\mu n_5\alpha_{[1}\beta_{2]}+\mu_5 n_5\alpha_{[1}\beta_{3]}\right)\Biggr\}$
		\\
		\hline
	\end{longtable}

	\bibliographystyle{utphys}
	\providecommand{\href}[2]{#2}\begingroup\raggedright\endgroup

\end{document}